\documentclass[12pt,dvipsnames]{article}
\usepackage[top=80pt,bottom=85pt,left=85pt,right=85pt]{geometry}

\pdfoutput=1
\usepackage[utf8]{inputenc}
\usepackage[T1]{fontenc} 
\usepackage[english]{babel} 
\usepackage{braket}
\usepackage{textcomp}
\usepackage[dvipsnames]{xcolor} %Package for fancy coloros
\usepackage{physics}
\usepackage{tabularx} 
\usepackage{ragged2e}
\usepackage{booktabs} 
\usepackage{rotating}
\usepackage{makecell,booktabs}
\usepackage{multirow} 
\usepackage{comment} 
\usepackage{amsmath}
\usepackage{cite}
\usepackage[all]{xy}

\usepackage{tabularray}

\usepackage{cite}

\usepackage{tikz}
\usetikzlibrary{math} 
\usetikzlibrary{3d} 
\usepackage{tikz-3dplot}
\usepackage{tikz-cd} 
\usepackage{microtype}
\usepackage{amsthm}
\usepackage{mathrsfs} 
\usepackage[strict]{changepage}
\newcommand\scalemath[2]{\scalebox{#1}{\mbox{\ensuremath{\displaystyle #2}}}}

\tikzcdset{
  cells={font=\everymath\expandafter{\the\everymath\displaystyle}},
}

\newcommand{\xdownarrow}[1]{%
  {\left\downarrow\vbox to #1{}\right.\kern-\nulldelimiterspace}
}

\usepackage{amssymb} 
\usepackage{bbold}
\usepackage{subcaption} 
\DeclareCaptionFormat{custom}

\usepackage[pdftex,colorlinks=true]{hyperref} 
\hypersetup{urlcolor=RedOrange, citecolor=burntorange, linkcolor=OliveGreen}

\usepackage{float} 
\usepackage{todonotes}

\usepackage{cleveref}
\makeatletter
\AddToHook{cmd/appendix/before}{\def\cref@section@alias{appendix}\def\cref@subsection@alias{appendix}}
\makeatother

\newcolumntype{E}{>{\hfil$}p{0.65cm}<{$\hfil}}

\newcolumntype{L}{>{\hfil$}p{16cm}<{$\hfil}}

\newcolumntype{D}{>{\hfil$}p{7.4cm}<{$\hfil}}
\newcolumntype{C}{>{\hfil$}p{3cm}<{$\hfil}}
\newcolumntype{P}{>{\hfil$}p{7.7cm}<{$\hfil}}
\newcolumntype{F}{>{\hfil$}p{5.7cm}<{$\hfil}}
\newcolumntype{S}{>{\hfil$}p{1.8cm}<{$\hfil}}
\newcolumntype{R}{>{\hfil$}p{5.2cm}<{$\hfil}}
\newcolumntype{U}{>{\hfil$}p{4.2cm}<{$\hfil}}
\newcolumntype{Q}{>{\hfil$}p{6.4cm}<{$\hfil}}
\newcolumntype{T}{>{\hfil$}p{1.9cm}<{$\hfil}}
\newcolumntype{V}{>{\hfil$}p{5.8cm}<{$\hfil}}
\newcolumntype{H}{>{\hfil$}p{1.8cm}<{$\hfil}}
\newcolumntype{A}{>{\hfil$}p{6cm}<{$\hfil}}
\newcolumntype{B}{>{\hfil$}p{2cm}<{$\hfil}}
\makeatletter
\newcommand\xleftrightarrow[2][]{%
  \ext@arrow 9999{\longleftrightarrowfill@}{#1}{#2}}
\newcommand\longleftrightarrowfill@{%
  \arrowfill@\leftarrow\relbar\rightarrow}
\makeatother

\newcommand{\C}{\mathbb{C}} 

\newcommand{\Z}{\mathbb{Z}}

\DeclareMathOperator{\Hom}{Hom}

\newtheorem{theorem}{Theorem}[section]

\newtheorem{remark}[theorem]{Remark}

\numberwithin{equation}{section}

\definecolor{burntorange}{rgb}{0.8, 0.33, 0.0}
\definecolor{cambridgeblue}{rgb}{0.64, 0.76, 0.68}
\definecolor{caribbeangreen}{rgb}{0.0, 0.8, 0.6}
\definecolor{celadon}{rgb}{0.67, 0.88, 0.69}
\definecolor{champagne}{rgb}{0.97, 0.91, 0.81}
\definecolor{cream}{rgb}{1.0, 0.99, 0.82}
\definecolor{cyan(process)}{rgb}{0.0, 0.72, 0.92}
\definecolor{brilliantlavender}{rgb}{0.96, 0.73, 1.0}
\definecolor{candypink}{rgb}{0.89, 0.44, 0.48}

\begin{document}

\begin{titlepage}

\phantom{wowiezowie}

\vspace{-1cm}

\begin{center}

{\Huge {\bf On the Orbifold origin of Higher Form Symmetries in Geometric Engineering}}

% \vspace{0.2cm}

% {\Huge {\bf A guide to McKay correspondance}}

%{\Huge {\bf The Case of Isometries}}

\vspace{1cm}

{\Large  Darius Dramburg$^*$, Shani Nadir Meynet$^{\ddagger}$ \\
and Andrea Sangiovanni$^{\sharp,\dagger,**}$}

\vspace{1cm}

{\it
{\small

{\footnotesize
$^*$ Kavli IPMU (WPI), UTIAS, The University of Tokyo, \\ Kashiwa, Chiba 277-8583, Japan \\ 
\vspace{0.25cm}
$^\sharp$ Mathematics Institute, Uppsala University, \\ Box 480, SE-75106 Uppsala, Sweden}\\
\vspace{.25cm}
{\footnotesize$^\dagger$ Department of Physics and Astronomy, Uppsala University,\\ Box 516, SE-75120 Uppsala, Sweden}\\
\vspace{.25cm}
{\footnotesize$^{**}$ Center for Geometry and Physics, Uppsala University,\\ Box 516, SE-75120 Uppsala, Sweden}\\
\vspace{.25cm}
$^\ddagger$ Department of Physics and Astronomy, University of Pennsylvania, \\ Philadelphia, PA 19104, USA
\vspace{.25cm}
\vspace{.25cm}
%$^*$ TODO\\
%\vspace{.25cm}
}}

\vskip -0.5cm
{\footnotesize \tt  darius.dramburg@ipmu.jp  \hspace{1cm} smeynet@sas.upenn.edu \hspace{1cm} andrea.sangiovanni@math.uu.se}

\vskip 1cm
     	{\bf Abstract }
\vskip .1in

\end{center}

\noindent In this work we explore the relation between orbifold singularities and higher form symmetries. Using the geometric engineering dictionary, we argue that the discrete higher symmetries of 5d SCFTs constructed from M-theory on a non-compact Calabi-Yau threefold can be related to a quantum symmetry of the associated BPS quiver. Through un-orbifolding the quantum symmetry we obtain a new theory without higher form symmetry, providing a notion of “minimality'' for a theory. This procedure is carried out via algebraic manipulations of the BPS/McKay quiver describing the crepant resolution of the singular geometry. This technique can also be reverted and thus, starting from any “minimal'' theory, one can orbifold it and generate new theories with the desired higher form symmetries. We test our technology on classes of 5d SCFTs that arise from M-theory geometric engineering on Calabi-Yau threefolds that are non-toric non-complete intersections, which have historically been challenging to tackle.

\end{titlepage}

% titlepage 
{
  \hypersetup{linkcolor=black}
  \tableofcontents
}
\newpage

\section{Introduction}

The seminal work by Gaiotto, Kapustin, Seiberg and Willett \cite{Gaiotto:2014kfa} sparked a renewed interest in the topic of symmetries in physics. In particular, symmetries of extended objects, higher form symmetries, have become a prominent research avenue both in the theoretical and phenomenological high energy community, see e.g.\ \cite{Sharpe:2015mja, McGreevy:2022oyu,
Freed:2022iao,
Gomes:2023ahz,
Schafer-Nameki:2023jdn,
Brennan:2023mmt,
Bhardwaj:2023kri,
Shao:2023gho,
Carqueville:2023jhb,
Costa:2024wks} for reviews. 

In the context of Geometric Engineering, higher form symmetries have been reliably studied in various setups, providing insight on the strong coupling regime of field theory. In this framework string/M-theory/F-theory is put on a non-compact CY/G$_2$ manifold background, used to engineer a field theory in the transverse space. Dynamical BPS particles of the theory are related to branes wrapping exceptional cycles in the resolved geometry, while extended rigid operators, the analogue of Wilson/'t Hooft lines, arise as branes wrapped on non-compact ones \cite{Katz:1996fh, Katz:1996th, Witten:1996qb, Morrison:1996xf, Intriligator:1997pq, Aharony:1997bh,Atiyah:2000zz, Acharya:2000gb, Heckman:2013pva, Jefferson:2018irk, Closset:2018bjz, Closset:2020scj, Apruzzi:2020zot, Closset:2020afy, Heidenreich:2020pkc, Bah:2020uev, Apruzzi:2021phx, Hosseini:2021ged, Apruzzi:2021vcu, Bhardwaj:2021wif, Bhardwaj:2021zrt, Closset:2021lwy, Heidenreich:2021xpr, Cvetic:2021maf, Debray:2021vob, Tian:2021cif, Braun:2021sex, Bah:2021brs, Cvetic:2021sxm, DelZotto:2022fnw, Cvetic:2022imb}.
't Hooft's screening argument, by which dynamical particles are able to break rigid line operators, can be translated into geometrical terms via the long exact sequence in homology
\begin{equation}\label{eq:long_sequence}
    \dots \rightarrow H_2(S^5/\Gamma) \overset{\imath_2}{\rightarrow} H_2(X/\Gamma) \overset{\jmath_2}{\rightarrow} H_2(X/\Gamma, S^5/\Gamma) \overset{\partial_2}{\rightarrow} H_1(S^5/\Gamma) \overset{\imath_1}{\rightarrow} H_1(X/\Gamma)\rightarrow \cdots,
\end{equation}
from which the so-called Defect Group\cite{DelZotto:2015isa,Albertini:2020mdx,Morrison:2020ool,Apruzzi:2021nmk, Baume:2023kkf} can be extracted:
\begin{align}\label{eq:defect_group}
    \mathbb{D} := \bigoplus_n \mathbb{D}^{(n)} \quad \text{where}\ \  \mathbb{D}^{(n)} =  \bigoplus_{p\text{-branes}}\left(\bigoplus_{k \text{ s.t. } \newline p-k+1=n} \left(\frac{H_k(\mathbf{X}, \partial \mathbf{X})}{H_{k}(\mathbf{X})}\right)\right) \, .
\end{align}

Physically, the defect group is the group acting on the unscreened lines of the field theory, while geometrically it describes torsional cycles in the boundary geometry. These two points of view come together in the linking picture described in \cite{Heckman:2022muc, Apruzzi:2022rei, DelZotto:2024tae, GarciaEtxebarria:2024fuk}.

Despite its simple description, the computation of the defect group is in general a hard task. However, in the special case of orbifold singularities, one can use the powerful result by Ito and Reid \cite{ItoReid} establishing a correspondence between the McKay quiver and the resolution of the singularity. The McKay quiver describes the quantum mechanics on the world volume of a D0 brane probing the singularity $\mathbb{C}^3/\Gamma$, and thus it can be identified with the BPS quiver of the transverse field theory, see \cite{Alim:2011kw, Closset:2019juk} for details. From this quiver it is then possible to easily read the defect group, a technique extensively used in (non-)supersymmetric theories \cite{DelZotto:2022fnw, Braeger:2024jcj, Braeger:2025rov}.

In this work we aim to extend the above result to orbifolds of base spaces which are more general than $\mathbb{C}^3$, in the context of M-theory geometric engineering on a non-compact Calabi-Yau threefold. Upon suitable conditions, this is thought to engineer a 5d SCFT on a flat spacetime background. \textit{Starting with a non-compact CY$_3$ $X$, for which the BPS quiver is known, we provide a general procedure to generate the quiver of the 5d SCFT engineered by the singularity $X/\Gamma$, for $\Gamma$ a finite subgroup of the automorphism group of $X$.} As the starting point of this program, we take 5d SCFTs for which the BPS quiver has been computed in the literature, or can be easily derived. The classes of threefolds $X$ that engineer such theories are substantially of three types:
\begin{itemize}
    \item toric singularities;
    \item compound Du Val (cDV) singularities;
    \item orbifolds $\mathbb{C}^3/\Gamma$, for $\Gamma \in SL(3,\mathbb{C})$.
\end{itemize}
Similar techniques were already used in the physics literature \cite{Feng:2000pc, Berenstein:2000mb, Garcia-Valdecasas:2019cqn}, but our task in this work is to show the way in which these results can be reinterpreted in the context of geometric engineering. This allows us to exhibit an explicit correspondence between the orbifold quantum symmetry and the higher form symmetries of the field theory. In particular, we prove that toric theories with higher form symmetries are always orbifolds of what we call “minimal'' theories, i.e.\ theories without higher form symmetries. Note that a “minimal'' theory can still be obtained via an orbifold of some other “minimal'' theory: our analysis aims only at relating theories with and without higher form symmetries via orbifolding.

As a welcome bonus side of our analysis, we show that the orbifolds of base spaces more general than $\mathbb{C}^3$ naturally give rise to a plethora of Calabi-Yau threefolds which are \textit{non-toric non-complete intersections}. These have been notoriously set aside in the geometric engineering literature, given that no clear mathematical handle is currently known to classify them, or to present their crepant resolution in full generality. We extract the singular threefolds for examples drawn from all the orbifold classes mentioned above. Thanks to the power of BPS quiver techniques, we provide essential data about the 5d SCFTs engineered by these threefolds, without the need to go through an explicit resolution procedure. Relatedly, crepant resolutions (as well as the corresponding BPS quivers) for examples in the class of orbifolds of cDV threefolds have been recently and carefully analyzed in \cite{Moleti}.

The paper is organized as follows: in \cref{sec:review} we review the basic aspects of 5d SCFTs and their BPS quivers with particular focus on the Defect Group. In section \cref{sec:orbifolding} we describe the orbifold procedure at the level of the BPS quiver with an explicit example. The formal derivation of the procedure can be found in \cref{app:skew_group}. \cref{sec:quantumdual} is devoted to the proof of the main claim, the correspondence between the Defect Group and the quantum symmetry of quiver. In \cref{sec:examples} a plethora of examples is presented, showing applications of the techniques presented in the previous sections.

\section{Defect Groups from BPS Quivers} \label{sec:review}

As discussed in the introduction, the Defect Group encodes the data of the spectrum of extended operators in a field theory. In the context of geometric engineering this can be read off from the intersection pairing among exceptional cycles in the engineering geometry \cite{DelZotto:2015isa,Albertini:2020mdx, Morrison:2020ool}. However, computing the intersection form is not always an easy task and it might rely on an explicit resolution of the singularities in the threefold, which are often technically challenging.\footnote{An alternative and equivalent approach is to compute the linking pairing between torsional 1- and 3-cycles at the base of the conical singularity, \cite{GarciaEtxebarria:2019caf, Apruzzi:2021nmk, Heckman:2022muc, DelZotto:2024tae}. In general this approach is no easier than computing a crepant resolution of the singular geometry.} However, thanks to the McKay correspondence, such tedious computation can be bypassed using quiver techniques. Indeed, it has been shown \cite{ItoReid, Ito1998McKayCA} that the McKay quiver associated to an orbifold encodes the data of the intersection form of the resolved singularity without the need to explicitly blown-up the geometry.

On the physical side, the McKay quiver is interpreted as the BPS quiver of the field theory transverse to the geometry, capturing the spectrum of its BPS particles. Exceptional cycles in the resolved geometry can be described as representations of the BPS quiver given suitable stability conditions \cite{Closset:2019juk}. Branes wrapping such cycles give rise to BPS particles in the field theory and given two particles, their mutual locality is governed by the Dirac pairing, encoded in the antisymmetric part of the adjacency matrix of the quiver. This matrix has been shown to encode the same data as the intersection pairing of the quiver \cite{Ito1998McKayCA}, and we will use it in the following to study the defect group of various examples. In this section we briefly review the crucial features of BPS quivers for 5d SCFTs and their relation to higher form symmetries.

\subsection{Quick review: 5d SCFTs and their BPS quiver}\label{sec: 5d SCFTs}
Our analysis is chiefly focused on extracting the 1-form symmetry of 5d SCFTs engineered by orbifold singularities via a powerfully indirect method, its associated BPS quiver, avoiding any explicit crepant resolution. Interestingly, the BPS quiver also contains information about the local degrees of freedom of the 5d SCFT, that we briefly review here. In this section we also recall some key facts about BPS quivers, the McKay correspondence and how they capture higher form symmetries. The expert reader can safely skip to the next sections for the new results.

An M-theory setup on a background of the form $X \times \mathbb{R}^{1,4}$, with $X$ a \textit{canonical} CY$_3$, is thought to engineer a 5d SCFT on the directions $\mathbb{R}^{1,4}$ transverse to the CY$_3$ \cite{Xie:2017pfl}. See also \cite{Closset:2020scj,Closset:2020afy,Closset:2021lwy,Bourget:2023wlb,Mu:2023uws,DeMarco:2023irn} for related work that employs this approach. The singularity is required to be canonical in order to ensure that it can be retrieved from a smooth CY$_3$ $\tilde{X}$ at finite distance in the moduli space of $\tilde{X}$, thus reaching the UV superconformal fixed point\footnote{Notice that $\tilde{X}$ is not guaranteed to exist, e.g.\ in the case of terminal singularities which do not admit small resolutions. Nonetheless, canonical singularities of this type are still well-defined, see e.g. \cite{Collinucci:2021wty,Collinucci:2021ofd,DeMarco:2022dgh} for more details about the associated 5d SCFTs.}. 

5d SCFTs are intrinsically non-Lagrangian, and can be partially characterized in terms of the following data:
\begin{itemize}
    \item The rank of the gauge symmetry $r$, which can be explicitly computed peforming a crepant resolution of $X$. In the language of homology:
    \begin{equation}
        r = \text{dim}H_4(\tilde{X},\mathbb{R}),
    \end{equation}
    with $\tilde{X}$ the resolved CY$_3$.
    \item The rank of the flavor symmetry, which can analogously be detected from the crepantly resolved phase $\tilde{X}$. In particular: 
     \begin{equation}
        f = \text{dim}H_2(\tilde{X},\mathbb{R})-\text{dim}H_4(\tilde{X},\mathbb{R}).
    \end{equation}
\end{itemize}

As we have mentioned in the Introduction, geometrical methods have been developed in order to compute the higher form symmetries of 5d SCFTs \cite{DelZotto:2015isa,Albertini:2020mdx,Morrison:2020ool}, as they do not admit a Lagrangian description, and are hence impenetrable for ordinary field-theoretic techniques. The simplest example of 5d SCFT with a higher form symmetry is Seiberg's $E_0$ theory \cite{Seiberg:1996bd}, which is realized via M-theory geometric engineering on the orbifold $\mathbb C^3 / \mathbb Z_3$ \cite{Morrison:1996xf}. The approach we will focus on in the rest on the paper relies on the connection between the BPS quiver and the intersection pairing of the engineering geometry.

From the field theory side, one obtains BPS states by wrapping exceptional cycles with branes. In the case of M-theory on CY$_3$, one can wrap M2 branes on 2-cycles, leading to point particles, and M5 branes on 4-cycles, leading to dynamical strings. Because of the electromagnetic pairing between M2 and M5 branes, the BPS particles and strings are in general mutually non-local, as electrons and magnetic monopoles in 4d QED. The non-locality is captured by the Dirac pairing between these states. As shown in \cite{Closset:2019juk}, 5d theories can as well be interpreted as 4d KK theories for which one can apply quiver techniques \cite{Alim:2011kw, Alim:2011ae}. We can now connect this field theory result to geometry via the McKay correspondence.

The striking result by McKay was to connect algebra and geometry via graph theory. Its original work relating the Dynkin graph of simply laced affine Lie algebras to the finite subgroup $\Gamma$ of $SU(2)$ and the resolution of the $\mathbb{C}^2/\Gamma$ orbifold allowed to rethink the problem of resolving a singularity not as task in algebra theory, but as a representation theory one \cite{MR604577, Gonzalez1983, Kronheimer1989TheCO}. The connection between McKay correspondence and field theory becomes manifest when considering the world volume theory of D0 branes probing a CY singularity. It turns out that the McKay graph can be interpreted as a $\mathcal{N}=4$ quantum mechanics on the D0 branes whose vacua are in one to one correspondence with the stable representations of the BPS quiver.

Given that the McKay quiver can be interpreted as the BPS quiver, it is now easy to describe the Defect Group of the theory encoding the higher form symmetries of the theory. Given two charge vectors $\gamma_i$ and $\gamma_j$, i.e.\ two choices of rank assignments for the BPS quiver, their Dirac pairing is given by the antisymmetric part of the BPS quiver's adjacency matrix:
\begin{align}\label{intersection matrix}
    B(\gamma_i, \gamma_j) = \gamma_i^t \cdot (A - A^t) \cdot \gamma_j \,.
\end{align}
Thus, one can interpret the Dirac pairing as a map between the charge lattice into itself $\gamma_i \in \mathbb{Z}^q \to B(.,\gamma_i) \in \mathbb{Z}^q$. We can now use 't Hooft's argument to conclude that the unscreened extended operators are precisely the ones with charge in the quotient lattice $\mathbb{Z}^q/B\mathbb{Z}^q$, \cite{Morrison:2020ool, Albertini:2020mdx, DelZotto:2022ras}. Thus, the Defect Group can be easily computed as the torsional part of the cokernel of the Dirac pairing. However, there is a subtlety concerning the equivalence between the Defect Group computed from geometry and the one from BPS quiver. It turns out \cite{Cvetic:2024dzu}, that the BPS defect group is in general a subgroup of the geometrical one, since the former is only sensible to the electro-magnetic pairing of the charges which translates to the linking pairing of the boundary geometry. Cycles with trivial pairing with any other cycle are effectively invisible to the BPS quiver analysis. In the rest of the paper, when we refer to Defect Group, we will considering the group effectively acting on the line defects of the theory.\footnote{We will comment on the geometric Defect Group in \cref{sec:conclusions} and how even the trivially acting part can in principle be detected by the BPS quiver, leaving a detailed analysis to future works.}

% The interested reader can check ... for further details on the relation between Dirac pairing, intersection pairing and topological operators in ZZZ, \cite{tHooft:1977nqb}, \cite{Caorsi:2017bnp}. 

Furthermore, essential data describing the 5d SCFT is encoded in the BPS quiver. The number of nodes $n$ in the quiver corresponds to:
 \begin{equation}
     n = 2r + f + 1,
 \end{equation}
 where “1'' accounts for the KK charge. Furthermore, the rank of the Dirac pairing matrix $B$ in \cref{intersection matrix} yields twice the gauge rank of the corresponding 5d SCFT:
 \begin{equation}
     \text{rank}(B) = 2r.
 \end{equation}
 Hence, knowledge of the quiver automatically encodes the data $r$ and $f$ of the 5d SCFT. This fact will be of much use for the remainder of this work, since it evades the technical challenges posed by performing an explicit crepant resolution of the threefold. In addition, the BPS quiver, via its representation theory, encodes the Gopakumar-Vafa invariants (or, equivalently, the Donaldson-Thomas invariants) of the associated Calabi-Yau threefold. For the scope of this work, we will not explore this avenue further.

\section{New quivers and back from orbifolding}\label{sec:orbifolding}

It is known \cite{Cohen:1984group, ReitenRiedtmann:1985, Garcia-Valdecasas:2019cqn, Berenstein:2000mb}, that from the McKay/BPS quiver associated to a certain singular geometry, a new one can obtained algorithmically whenever there is a group action on the quiver itself. This new quiver can be interpreted again as the BPS quiver of new theory, as proven in \cite{Bridgeland:2024saj}, and the new quiver turns out to be the one associated to an orbifolding of the starting geometry by the same group.

The scope of this section is to review this procedure of (un-)orbifolding quivers to describe how higher form symmetries arises in field theory. When the theory of interest is described by a geometry with a $\mathbb{C}^*$ action, the orbifold procedure is particularly easy, as we will show momentarily with an explicit example, leaving the general derivation in \cref{app:skew_group}, where the mathematical details are reviewed.

\subsection{Quiver orbifold: big from small}

In this section we consider a specific example of a general technique described in \cref{app:skew_group}. The aim is to familiarize the reader with this procedure that will then be used to prove the general result of \cref{sec:quantumdual}, and applied extensively in \cref{sec:examples}.

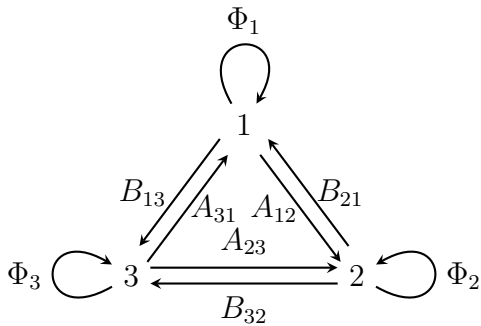
\begin{figure}[H]
\begin{center}
\begin{tikzpicture}[>=stealth, thick]
% Nodes
\node (1) at (0,0) {$1$};
\node (2) at (1.5,-2) {$2$};
\node (3) at (-1.5,-2) {$3$};
% Loops
\draw[->] (1) edge[->, min distance=12mm, in=60, out=120] node[above]{$\Phi_1$} (1);
\draw[->] (2) edge[->, min distance=12mm, in=30, out=-30] node[right]{$\Phi_2$} (2);
\draw[->] (3) edge[->, min distance=12mm, in=150, out=-150] node[left]{$\Phi_3$} (3);
%%%%%
\draw[<-, transform canvas={xshift=3pt,yshift=3pt}] (1)--(2)   node[midway,xshift=12pt] {$B_{21}$};
\draw[<-, transform canvas={yshift=-3pt}] (2)--(1) node[midway,xshift=-10pt] {$A_{12}$};
\draw[->, transform canvas={xshift=-3pt,yshift=3pt}](1)--(3)node[midway,xshift=-14pt] {$B_{13}$};
\draw[->, transform canvas={yshift=-3pt}] (3)--(1)node[midway,xshift=10pt] {$A_{31}$};
\draw[<-, transform canvas={yshift=3pt}] (2)--(3)node[midway,yshift=10pt] {$A_{23}$};
\draw[<-, transform canvas={yshift=-3pt}] (3)--(2)node[midway,yshift=-10pt] {$B_{32}$};
\end{tikzpicture}
\end{center}
    \caption{BPS quiver for $\mathbb{C}^2/\mathbb{Z}_3\times \mathbb{C}$.}
    \label{fig:placeholder}
\end{figure}

%\begin{figure}[H]
%    \centering
 %   \includegraphics[width=0.25\linewidth]{c3:z3.png}
%    \caption{Caption}
%    \label{fig:placeholder}
%\end{figure}
We use as a working example the quiver associated to the geometry $\mathbb{C}^2/\mathbb{Z}_3\times \mathbb{C}$ given in \Cref{fig:placeholder}. This orbifold can be equivalently described via the algebraic equation $xy=z^3$ in $\mathbb{C}^4$. If one interprets the above quiver as the QFT living on a D3 probing the orbifold, the algebraic equation can be recovered as the equation describing the moduli space of the theory. Let us see review this explicitly. If we assign a $U(1)$ gauge group to each node, we have the following gauge invariants fields
\begin{align}
    X = A_{12}A_{23}A_{31} \,,  \quad Y = A_{13}A_{32}A_{21} \,, \quad Z_{ij} = A_{ij}A_{ji} \,, \quad W_i=\Phi_i \, .
\end{align}
Imposing the F-term equations we get:
\begin{align}
    Z_{ij} = Z \,, \quad W_i=W \, , \quad XY=Z^3 \, ,
\end{align}
in accordance with the algebraic equation.

We can now consider the further orbifold $(\mathbb{C}^2/\mathbb{Z}_3\times \mathbb{C})/\mathbb{Z}_3$ with the action
\begin{align}
    (x,y,z,w) \to (\omega \, x, \omega^2 \, y, \omega \, z, \omega^2 \, w) \, , \quad \omega^3=1.
\end{align}
The new BPS quiver can be computed from the previous one considering the $\mathbb{Z}_3$ action on the arrows $A_{ij}, \Phi_i$ that reproduces the above action on the gauge invariant coordinates. In the case at hand we have
\begin{align}\label{eq:field_action_z3}
    A_{12} \to \omega \, A_{12} \, , \quad  A_{32} \to \omega \, A_{32} \, , \quad  A_{13} \to \omega \, A_{13} \, , \quad  \Phi_{i} \to \omega^2 \, \Phi_{i} \, ,
\end{align}
while the action on all other arrows is trivial.

The quiver for the $(\mathbb{C}^2/\mathbb{Z}_3\times \mathbb{C})/\mathbb{Z}_3$ theory is now given by applying the procedure detailed in \cref{app:skew_group}.\footnote{For the Chan-Paton analysis of the example we refer to \cref{app:chan_paton}.} Each node of the parent quiver splits into $n$ nodes, where $n$ is the number of irreps of the orbifolding group and each arrow connecting nodes $i$ and $j$ becomes an arrow connecting irreps $\rho_i$ of the $i$ node with irreps $\tau_j$ of the $j$ node if $Hom(\rho, \tau \otimes R(A_{ij}))$ is non-zero. 

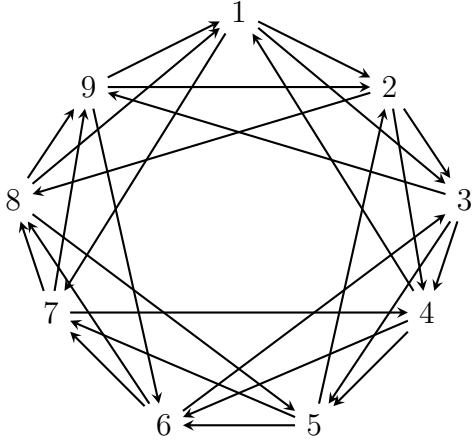
\begin{figure}[H]
\begin{center}
\begin{tikzpicture}[>=stealth, thick]
% Nodes
\node (1) at (0,0) {$1$};
\node (2) at (2,-1) {$2$};
\node (3) at (3,-2.5) {$3$};
\node (4) at (2.5,-4) {$4$};
\node (5) at (1,-5.5) {$5$};
\node (6) at (-1,-5.5) {$6$};
\node (7) at (-2.5,-4) {$7$};
\node (8) at (-3,-2.5) {$8$};
\node (9) at (-2,-1) {$9$};
% R → M arrows (upper and lower)
\draw[->] (1) to node[above]{} (2);
\draw[->] (2) to node[above]{} (3);
\draw[->] (3) to node[above]{} (4);
\draw[->] (4) to node[above]{} (5);
\draw[->] (5) to node[above]{} (6);
\draw[->] (6) to node[above]{} (7);
\draw[->] (7) to node[above]{} (8);
\draw[->] (8) to node[above]{} (9);
\draw[->] (9) to node[above]{} (1);
%%%%
\draw[->] (1) to node[above]{} (3);
\draw[->] (1) to node[above]{} (7);
\draw[->] (2) to node[above]{} (4);
\draw[->] (2) to node[above]{} (8);
\draw[->] (3) to node[above]{} (5);
\draw[->] (3) to node[above]{} (9);
\draw[->] (4) to node[above]{} (1);
\draw[->] (4) to node[above]{} (6);
\draw[->] (5) to node[above]{} (2);
\draw[->] (5) to node[above]{} (7);
\draw[->] (6) to node[above]{} (3);
\draw[->] (6) to node[above]{} (8);
\draw[->] (7) to node[above]{} (4);
\draw[->] (7) to node[above]{} (9);
\draw[->] (8) to node[above]{} (5);
\draw[->] (8) to node[above]{} (1);
\draw[->] (9) to node[above]{} (2);
\draw[->] (9) to node[above]{} (6);
%%%%
\end{tikzpicture}\\
\end{center}
\caption{BPS quiver for $\mathbb{C}^3/\mathbb{Z}_9$ with action $(1,2,6)$.}
\label{fig:placeholder2}
\end{figure}

In our running example we have that each node $i$ splits into three nodes $i_{1,\omega,\omega^2}$, labeled by irreps of $\mathbb{Z}_3$ and the arrows $A_{ij}$ will now connect node $i_{1,\omega,\omega^2}$ to node $j_{1,\omega,\omega^2}$ depending on the action of the representation they transform in \cref{eq:field_action_z3}. For example $A_{12}$ transforms in the $\omega$ irrep of $\mathbb{Z}_3$, thus it  will connect nodes $1_{1}$ and $2_{\omega^2}$, since $Hom(\rho, \tau \otimes R(A_{ij})) = Hom(1, \tau \otimes \omega) = 1 \iff \tau = \omega^2 $. Repeating this procedure for each arrow and node we get the new quiver, depicted in Figure \ref{fig:placeholder2}.

%\begin{figure}[H]
%    \centering
%    \includegraphics[width=0.25\linewidth]{c3:z9.png}
%   \caption{Caption}
%   \label{fig:placeholder2}
%end{figure}

Clearly, Figure \ref{fig:placeholder2} turns out to be the quiver for $\mathbb{C}^3/\mathbb{Z}_9$ with action $(1,2,6)$. As proven in detail in \cref{app:skew_group}, this procedure can be applied to any quiver and doesn't rely on the fact that we started with a quiver associated to an orbifold of $\mathbb{C}^3$.

Before moving to the next section, let us compute the defect group of this theory. The torsional part of the cokernel of the Dirac pairing can be computed via the Smith algorithm to be $\mathbb{Z}_3^2$: thus the theory admits a $\mathbb{Z}_3$ electric and magnetic center symmetries. This symmetry is $\mathbb{Z}_3$ not by coincidence, as we will show later, but as a consequence of the orbifolding procedure.

\subsection{Quiver unorbifold: small from big}

The procedure outlined in the previous section can be reverted, so that given a quiver with a symmetry, one can obtain a new quiver identifying nodes and arrows on the same orbit of the symmetry action. 

Let us go back to the running example of the $\mathbb{Z}_9$ orbifold described previously. The quiver admits a symmetry $\mathbb{Z}_3$ mapping node $i_\rho$ to node $i_{\omega \otimes \rho}$, for $\omega$ an irrep of $\mathbb{Z}_3$. If we identify all the nodes and arrows in the $\mathbb{Z}_3$ action, we get back the quiver of $\mathbb{C}^2/\mathbb{Z}_3 \times \mathbb{C}$. This fact can be explained both via physical and mathematical arguments.

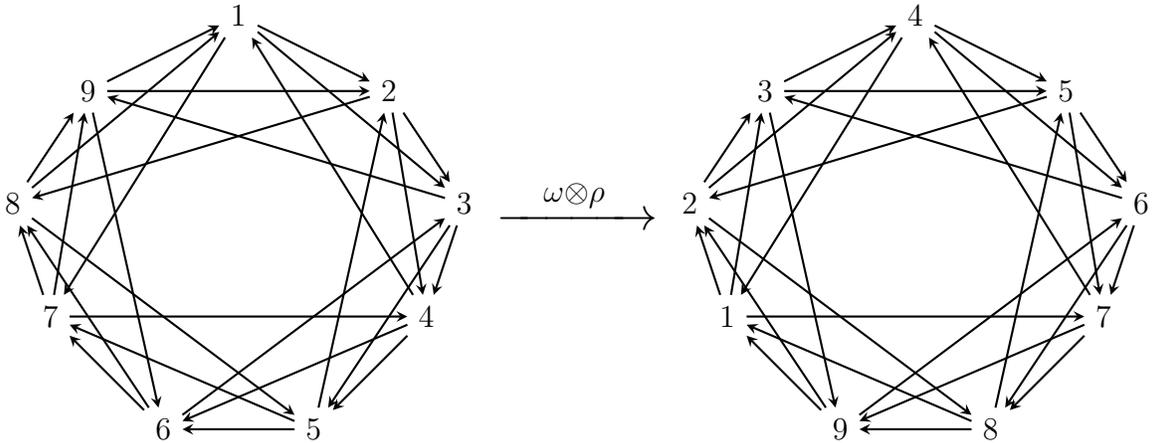
\begin{figure}[H]
\begin{tikzpicture}[>=stealth, thick]
% Nodes
\node (1) at (0,0) {$1$};
\node (2) at (2,-1) {$2$};
\node (3) at (3,-2.5) {$3$};
\node (4) at (2.5,-4) {$4$};
\node (5) at (1,-5.5) {$5$};
\node (6) at (-1,-5.5) {$6$};
\node (7) at (-2.5,-4) {$7$};
\node (8) at (-3,-2.5) {$8$};
\node (9) at (-2,-1) {$9$};
% R → M arrows (upper and lower)
\draw[->] (1) to node[above]{} (2);
\draw[->] (2) to node[above]{} (3);
\draw[->] (3) to node[above]{} (4);
\draw[->] (4) to node[above]{} (5);
\draw[->] (5) to node[above]{} (6);
\draw[->] (6) to node[above]{} (7);
\draw[->] (7) to node[above]{} (8);
\draw[->] (8) to node[above]{} (9);
\draw[->] (9) to node[above]{} (1);
%%%%
\draw[->] (1) to node[above]{} (3);
\draw[->] (1) to node[above]{} (7);
\draw[->] (2) to node[above]{} (4);
\draw[->] (2) to node[above]{} (8);
\draw[->] (3) to node[above]{} (5);
\draw[->] (3) to node[above]{} (9);
\draw[->] (4) to node[above]{} (1);
\draw[->] (4) to node[above]{} (6);
\draw[->] (5) to node[above]{} (2);
\draw[->] (5) to node[above]{} (7);
\draw[->] (6) to node[above]{} (3);
\draw[->] (6) to node[above]{} (8);
\draw[->] (7) to node[above]{} (4);
\draw[->] (7) to node[above]{} (9);
\draw[->] (8) to node[above]{} (5);
\draw[->] (8) to node[above]{} (1);
\draw[->] (9) to node[above]{} (2);
\draw[->] (9) to node[above]{} (6);
%%%%

\node (1) at (4.5,-2.5) {\Large $\xrightarrow{\hspace{0.5cm}\omega \otimes \rho\hspace{0.5cm}}$}; 

%%%%%
% Nodes
\node (1) at (9,0) {$4$};
\node (2) at (11,-1) {$5$};
\node (3) at (12,-2.5) {$6$};
\node (4) at (11.5,-4) {$7$};
\node (5) at (10,-5.5) {$8$};
\node (6) at (8,-5.5) {$9$};
\node (7) at (6.5,-4) {$1$};
\node (8) at (6,-2.5) {$2$};
\node (9) at (7,-1) {$3$};
% R → M arrows (upper and lower)
\draw[->] (1) to node[above]{} (2);
\draw[->] (2) to node[above]{} (3);
\draw[->] (3) to node[above]{} (4);
\draw[->] (4) to node[above]{} (5);
\draw[->] (5) to node[above]{} (6);
\draw[->] (6) to node[above]{} (7);
\draw[->] (7) to node[above]{} (8);
\draw[->] (8) to node[above]{} (9);
\draw[->] (9) to node[above]{} (1);
%%%%
\draw[->] (1) to node[above]{} (3);
\draw[->] (1) to node[above]{} (7);
\draw[->] (2) to node[above]{} (4);
\draw[->] (2) to node[above]{} (8);
\draw[->] (3) to node[above]{} (5);
\draw[->] (3) to node[above]{} (9);
\draw[->] (4) to node[above]{} (1);
\draw[->] (4) to node[above]{} (6);
\draw[->] (5) to node[above]{} (2);
\draw[->] (5) to node[above]{} (7);
\draw[->] (6) to node[above]{} (3);
\draw[->] (6) to node[above]{} (8);
\draw[->] (7) to node[above]{} (4);
\draw[->] (7) to node[above]{} (9);
\draw[->] (8) to node[above]{} (5);
\draw[->] (8) to node[above]{} (1);
\draw[->] (9) to node[above]{} (2);
\draw[->] (9) to node[above]{} (6);
\end{tikzpicture}\\
\caption{The rotation action identifying nodes 147, 258 and 369.}
\label{fig:placeholder3}
\end{figure}

%\begin{figure}[H]
%    \centering
%    \includegraphics[width=1\linewidth]{rotatz3.png}
%    \caption{The rotation action identifying nodes 147, 258 and 369.}
%    \label{fig:placeholder}
%\end{figure}

From the physical point of view, this amounts to nothing but the gauging of the “quantum symmetry'' that arises after orbifolding. This fact is well known in the CFT context, where gauging an abelian group $\mathbb{Z}_n$ and then gauging the quantum dual $\widehat{ \mathbb{Z}}_n$\footnote{We return in more detail on the role of the quantum dual symmetry in \cref{sec:quantumdual}.}, that acts on the twisted sectors, gives back the original theory. Indeed, from the point of view of the worldsheet theory of a string, we are going from a theory with target $\mathbb{C}^2/\mathbb{Z}_3\times \mathbb{C}$ to another with target $(\mathbb{C}^2/\mathbb{Z}_3\times \mathbb{C})/\mathbb{Z}_3$ and back.

From the mathematical point of view, the un-orbifolding procedure can be described as a skewing of the path algebra of the quiver. The technical definition is beyond the scope of this section and the interested reader can check \cref{app:skew_group} for further details.

This observation concludes the (un-)orbifolding procedure on the quiver. In what follows we will describe how the “quantum symmetry'' that arises after orbifolding can be directly related to the higher form symmetries of the theory associated to the quiver. In particular, in the case of toric geometries, we show that any theory admitting a higher form symmetry, can be obtained as the orbifold of another theory, free of higher form symmetries. Let us note that, in general, orbifolding may not introduce higher form symmetries on the new theory. However, as we will discuss in the next section, the orbifolding procedure that generates a higher form symmetry are precisely the ones that act without fixed loci on the starting geometry.

\section{Quantum dual and higher form symmetry}\label{sec:quantumdual}

The connection between higher form symmetries and orbifolds was hinted at in \cite{DelZotto:2022fnw}. Using the classic results by Armstrong \cite{armstrong1968fundamental}:

\textit{Let $\Gamma$ be a discontinuous group of homeomorphisms of a path connected, simply connected, locally compact metric space $SE$, and let $H$ be the normal subgroup of $\Gamma$ generated by those elements which have fixed points. Then the fundamental group of the orbit space $SE/\Gamma$ is isomorphic to the factor group $\Gamma/H$.} 

This theorem can be used to compute the Defect Group from the action of an orbifold on the boundary space of the engineering geometry. In this section we revisit this fact employing quiver techniques, bypassing the need to know the actual algebraic description of the geometry.

The main result of this section is that the Defect Group can be put in correspondence with the so called “quantum symmetry'' of the quiver, a special symmetry that permutes the nodes of a quiver. Starting with the simple example of $\mathbb{C}^3/\Gamma$ singularities, for $\Gamma$ abelian, we extend this result to any toric singularity. This equivalence allows us to define “minimal'' theories, i.e.\ any theory without higher form symmetries, since any other theory with such symmetries can be obtained as an orbifold of a minimal one.

\subsection{Abelian orbifolds of $\mathbb{C}^3$}

The crucial ingredient to establish the higher form symmetry - orbifold correspondence is the notion of “quantum symmetry'' anticipated in the previous section. The McKay quiver constructed from an abelian $\mathbb{C}^3/\Gamma$ orbifold comes with a natural symmetry on it, given by tensoring the representation associated to each node with another 1d irrep of the orbifold group. This action induces a $\Gamma$ symmetry, often dubbed $\widehat \Gamma$, called the “quantum symmetry'' or “quantum dual'' symmetry.\footnote{See section 5.6 of \cite{Aspinwall:2009isa}.}

\begin{figure}[H]
\begin{tikzpicture}[>=stealth, thick]
% Nodes
\node (1) at (0,0) {$1$};
\node (2) at (2,-1) {$2$};
\node (3) at (3,-2.5) {$3$};
\node (4) at (2.5,-4) {$4$};
\node (5) at (1,-5.5) {$5$};
\node (6) at (-1,-5.5) {$6$};
\node (7) at (-2.5,-4) {$7$};
\node (8) at (-3,-2.5) {$8$};
\node (9) at (-2,-1) {$9$};
% R → M arrows (upper and lower)
\draw[->] (1) to node[above]{} (2);
\draw[->] (2) to node[above]{} (3);
\draw[->] (3) to node[above]{} (4);
\draw[->] (4) to node[above]{} (5);
\draw[->] (5) to node[above]{} (6);
\draw[->] (6) to node[above]{} (7);
\draw[->] (7) to node[above]{} (8);
\draw[->] (8) to node[above]{} (9);
\draw[->] (9) to node[above]{} (1);
%%%%
\draw[->] (1) to node[above]{} (3);
\draw[->] (1) to node[above]{} (7);
\draw[->] (2) to node[above]{} (4);
\draw[->] (2) to node[above]{} (8);
\draw[->] (3) to node[above]{} (5);
\draw[->] (3) to node[above]{} (9);
\draw[->] (4) to node[above]{} (1);
\draw[->] (4) to node[above]{} (6);
\draw[->] (5) to node[above]{} (2);
\draw[->] (5) to node[above]{} (7);
\draw[->] (6) to node[above]{} (3);
\draw[->] (6) to node[above]{} (8);
\draw[->] (7) to node[above]{} (4);
\draw[->] (7) to node[above]{} (9);
\draw[->] (8) to node[above]{} (5);
\draw[->] (8) to node[above]{} (1);
\draw[->] (9) to node[above]{} (2);
\draw[->] (9) to node[above]{} (6);
%%%%

\node (1) at (4.5,-2.5) {\Large $\xrightarrow{\hspace{0.5cm}\rho_2 \otimes \rho_i\hspace{0.5cm}}$}; 

%%%%%
% Nodes
\node (1) at (9,0) {$2$};
\node (2) at (11,-1) {$3$};
\node (3) at (12,-2.5) {$4$};
\node (4) at (11.5,-4) {$5$};
\node (5) at (10,-5.5) {$6$};
\node (6) at (8,-5.5) {$7$};
\node (7) at (6.5,-4) {$8$};
\node (8) at (6,-2.5) {$9$};
\node (9) at (7,-1) {$1$};
% R → M arrows (upper and lower)
\draw[->] (1) to node[above]{} (2);
\draw[->] (2) to node[above]{} (3);
\draw[->] (3) to node[above]{} (4);
\draw[->] (4) to node[above]{} (5);
\draw[->] (5) to node[above]{} (6);
\draw[->] (6) to node[above]{} (7);
\draw[->] (7) to node[above]{} (8);
\draw[->] (8) to node[above]{} (9);
\draw[->] (9) to node[above]{} (1);
%%%%
\draw[->] (1) to node[above]{} (3);
\draw[->] (1) to node[above]{} (7);
\draw[->] (2) to node[above]{} (4);
\draw[->] (2) to node[above]{} (8);
\draw[->] (3) to node[above]{} (5);
\draw[->] (3) to node[above]{} (9);
\draw[->] (4) to node[above]{} (1);
\draw[->] (4) to node[above]{} (6);
\draw[->] (5) to node[above]{} (2);
\draw[->] (5) to node[above]{} (7);
\draw[->] (6) to node[above]{} (3);
\draw[->] (6) to node[above]{} (8);
\draw[->] (7) to node[above]{} (4);
\draw[->] (7) to node[above]{} (9);
\draw[->] (8) to node[above]{} (5);
\draw[->] (8) to node[above]{} (1);
\draw[->] (9) to node[above]{} (2);
\draw[->] (9) to node[above]{} (6);
\end{tikzpicture}\\
\caption{The rotation action identifying nodes 147, 258 and 369.}
\label{fig:placeholder4}
\end{figure}
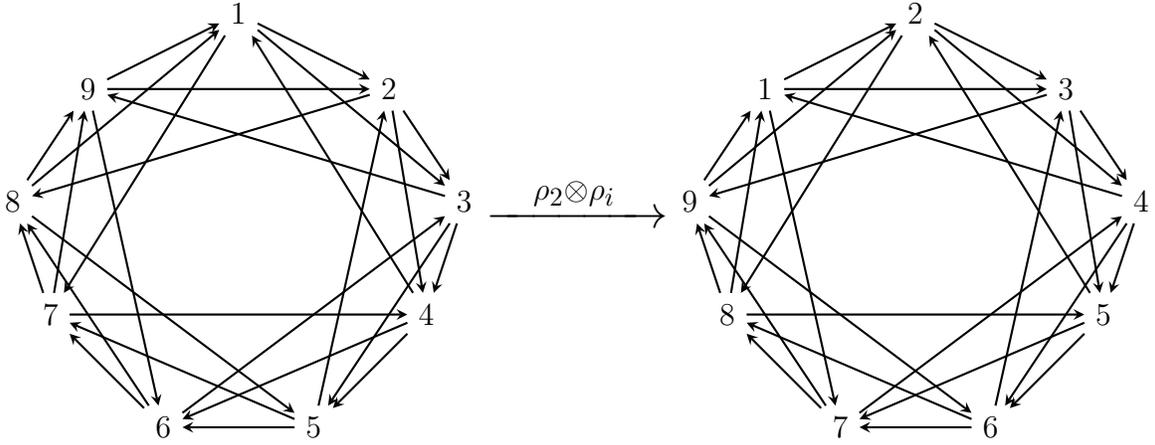

Let us look at the example $\mathbb{C}^3/\mathbb{Z}_9$ discussed in the previous sections. This theory admits a $\widehat{\mathbb{Z}}_9$ quantum symmetry given by clockwise rotations of the quiver in \Cref{fig:placeholder4}, similar to the one discussed in the previous section.

In order to establish the correspondence between the the quantum symmetry with the higher form symmetry, we need to look at how the orbifold is acting. The action 
\begin{equation}
   (1,2,6) \to g\cdot(1,2,6), \quad \text{with } g = diag(\omega,\omega^2,\omega^6),
\end{equation}
 has fixed loci for $g^0,g^3,g^6$. Looking at the character table of the group, we see that the irreps corresponding to $\omega^{0,3,6}$ are precisely the ones that trivialize the action of $g^0,g^3,g^6$.\footnote{This can be seen from the fact that $\chi_{\omega}(g)=1$.} From the analysis performed in the appendix of \cite{DelZotto:2022fnw}, we know that the defect group can be read as the abelianization of the orbifold subgroup that acts without fixed loci. This subgroup can be obtained as Ab$(\Gamma/N)$, where $N$ can be read from the character table of the group $\Gamma$ by taking the intersection of all conjugacy classes of $\Gamma$ in the kernel of the representation by which the orbifold acts on $\mathbb{C}^3$.\footnote{In other words, given a representation $\rho$ of $\Gamma$, $N$ is given by the intersection of all conjugacy classes for which $\chi_{\rho}(g) = \chi_{\rho}(1)$, i.e.\ the identity element.} In the case at hand the character table is given in \cref{eq:chatbl} and the representation acting on $\C^3$ is $R = \rho_2 \oplus \rho_3 \oplus \rho_7$.
 \begin{align}\label{eq:chatbl}
 \begin{matrix}
    \rho_1 \\
    \rho_2 \\
    \rho_3 \\
    \rho_4 \\
    \rho_5 \\
    \rho_6 \\
    \rho_7 \\
    \rho_8 \\
    \rho_9 \\
\end{matrix}
\begin{pmatrix}
1 & 1 & 1 & 1 & 1 & 1 & 1 & 1 & 1  \\
1 & \zeta & \zeta^2 & \zeta^3 & \zeta^4 & \zeta^5 & \zeta^6 & \zeta^7 & \zeta^8 \\
1 & \zeta^2 & \zeta^4 & \zeta^2 & \zeta^8 & \zeta^1 & \zeta^3 & \zeta^5 & \zeta^7 \\
1 & \zeta^3 & \zeta^6 & 1 & \zeta^3 & \zeta^6 & 1 & \zeta^3 & \zeta^6 \\
1 & \zeta^4 & \zeta^8 & \zeta^3 & \zeta^7 & \zeta^2 & \zeta^6 & \zeta^1 & \zeta^5 \\
1 & \zeta^5 & \zeta^1 & \zeta^6 & \zeta^2 & \zeta^7 & \zeta^3 & \zeta^8 & \zeta^4 \\
1 & \zeta^6 & \zeta^3 & 1 & \zeta^6 & \zeta^3 & 1 & \zeta^6 & \zeta^3 \\
1 & \zeta^7 & \zeta^5 & \zeta^3 & \zeta^1 & \zeta^8 & \zeta^6 & \zeta^4 & \zeta^2 \\
1 & \zeta^8 & \zeta^7 & \zeta^6 & \zeta^5 & \zeta^4 & \zeta^3 & \zeta^2 & \zeta^1 
\end{pmatrix}
\end{align}
 
 We thus have that the center symmetry can be read from the exact sequence
\begin{align}\label{eq:exact_quantum_group}
  0 \to  N \to \Gamma \to \Gamma/N \to 0 \, .
\end{align}
Finally, we can associate to each representation of $\Gamma/N$ its induced representation in $\Gamma$, and thus a subgroup of the quantum dual. It turns out that this subgroup is given precisely by the irreps we identified above: they are the ones that trivialize the group elements that act with fixed loci.

This brings us to the main result of this section:

\textit{We have established a correspondence between the defect group and the quantum symmetry acting on the quiver. By the procedure outlined in the previous section we can now un-orbifold this action, leading to a $\mathbb{C}^3/N$ orbifold, which by construction will not have a center symmetry.}

\begin{figure}[H]
 \begin{tikzpicture}
% Nodes
\node (1) at (0,0) {$1$};
\node (2) at (2,-1) {$2$};
\node (3) at (3,-2.5) {$3$};
\node (4) at (2.5,-4) {$4$};
\node (5) at (1,-5.5) {$5$};
\node (6) at (-1,-5.5) {$6$};
\node (7) at (-2.5,-4) {$7$};
\node (8) at (-3,-2.5) {$8$};
\node (9) at (-2,-1) {$9$};
% R → M arrows (upper and lower)
\draw[->] (1) to node[above]{} (2);
\draw[->] (2) to node[above]{} (3);
\draw[->] (3) to node[above]{} (4);
\draw[->] (4) to node[above]{} (5);
\draw[->] (5) to node[above]{} (6);
\draw[->] (6) to node[above]{} (7);
\draw[->] (7) to node[above]{} (8);
\draw[->] (8) to node[above]{} (9);
\draw[->] (9) to node[above]{} (1);
%%%%
\draw[->] (1) to node[above]{} (3);
\draw[->] (1) to node[above]{} (7);
\draw[->] (2) to node[above]{} (4);
\draw[->] (2) to node[above]{} (8);
\draw[->] (3) to node[above]{} (5);
\draw[->] (3) to node[above]{} (9);
\draw[->] (4) to node[above]{} (1);
\draw[->] (4) to node[above]{} (6);
\draw[->] (5) to node[above]{} (2);
\draw[->] (5) to node[above]{} (7);
\draw[->] (6) to node[above]{} (3);
\draw[->] (6) to node[above]{} (8);
\draw[->] (7) to node[above]{} (4);
\draw[->] (7) to node[above]{} (9);
\draw[->] (8) to node[above]{} (5);
\draw[->] (8) to node[above]{} (1);
\draw[->] (9) to node[above]{} (2);
\draw[->] (9) to node[above]{} (6);
%%%%

\node (1) at (4.5,-2.5) {$\xrightarrow{\hspace{1.5cm}\hspace{0.5cm}}$}; 

%%%%%
 \scalemath{0.8}{
     % Nodes
\node (1) at (11.5,-1.5) {$(1,4,7)$};
\node (2) at (13.5,-4.5) {$(2,5,8)$};
\node (3) at (9.5,-4.5) {$(3,6,9)$};
% Loops
\draw[->] (1) edge[->, min distance=12mm, in=60, out=120] node[above]{\footnotesize$A_{17}\sim A_{41}\sim A_{74}$} (1);
\draw[->] (2) edge[->, min distance=12mm, in=30, out=-30] node[right]{\footnotesize$A_{52}\sim A_{28}\sim A_{85}$} (2);
\draw[->] (3) edge[->, min distance=12mm, in=150, out=-150] node[left]{\footnotesize$A_{39}\sim A_{96}\sim A_{63}$} (3);
%%%%%
\draw[<-, transform canvas={xshift=3pt,yshift=3pt}] (1)--(2)   node[midway,xshift=48pt] {$A_{12}\sim A_{45}\sim A_{78} $};
\draw[<-, transform canvas={yshift=-3pt}] (2)--(1) node[midway,xshift=-10pt] {};
\draw[->, transform canvas={xshift=-3pt,yshift=3pt}](1)--(3)node[midway,xshift=-14pt] {};
\draw[->, transform canvas={yshift=-3pt}] (3)--(1)node[midway,xshift=10pt] {};
\draw[<-, transform canvas={yshift=3pt}] (2)--(3)node[midway,yshift=10pt] {};
\draw[<-, transform canvas={yshift=-3pt}] (3)--(2)node[midway,yshift=-10pt] {};
}
\end{tikzpicture}
 \caption{Unorbifolding procedure that produces a theory with no defect group. The orbifold action identifies the nodes on the left-hand side. E.g.\ the nodes $(1,4,7)$ are identified by the action, and appear as a single node on the right-hand side. The same goes for arrows: we denote as $A_{ij}$ an arrow from node $i$ to node $j$ in the quiver on the left. We explicitly show some identified arrows in the quiver on the right.}
 \label{fig: Z3 quotient}
\end{figure}
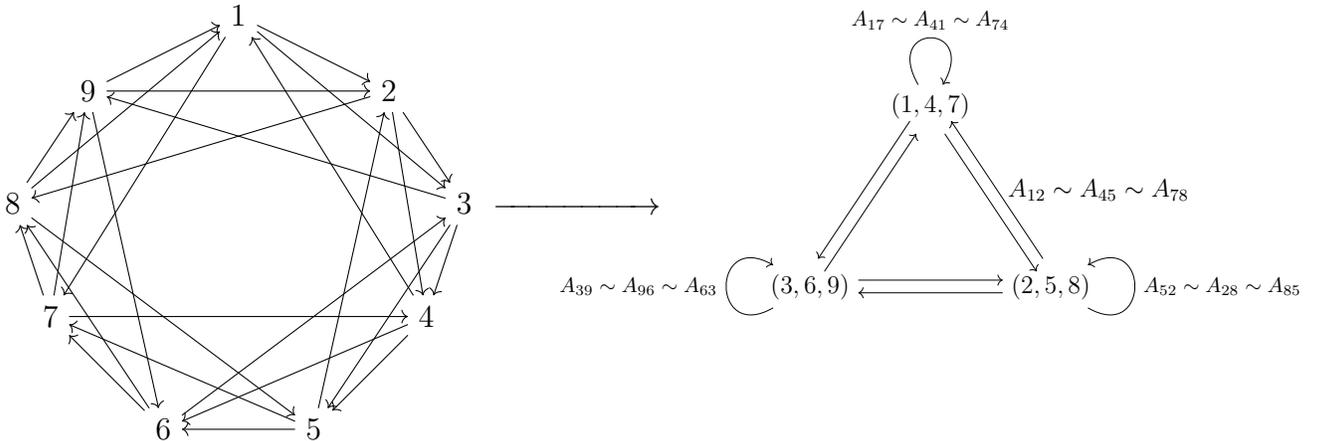

This can be checked explicitly by quotienting the quantum subgroup just described. By applying the prescription of \cref{app:skew_group} we obtain the quiver on the right-hand side of Figure \ref{fig: Z3 quotient}, which is again $\mathbb{C}^2/\mathbb{Z}_3\times \mathbb{C}$. 

From this analysis it is now obvious that for any theory engineered from $\mathbb{C}^3/(\mathbb{Z}_n\times \mathbb{Z}_m)$ one can un-orbifold the quantum symmetry associated to the higher form one, $\widehat{ \mathbb{Z}}_k$, to get a theory engineered by $(\mathbb{C}^3/(\mathbb{Z}_n\times \mathbb{Z}_m))/\widehat{ \mathbb{Z}}_k$. More generally, one can fully un-orbifold the quiver, leading to the $\C^3$ geometry, which in this case leads to the empty theory. 

The process just described can be extended beyond abelian orbifolds of $\mathbb{C}^3$, in order to encompass any toric geometry, as we discuss in the next section.

\subsection{The toric case in general}

In order to extend the previous analysis to a generic toric threefold we use the result of \cite{Garcia-Valdecasas:2019cqn}. We do not review in depth the dimer/toric correspondence (for additional details we refer to \cite{Hanany:2005ve, Franco:2005rj}), but we use the fact that the toric diagram contains all the data of the quantum symmetry, even if the quiver is not explicitly constructed.

The result of \cite{Garcia-Valdecasas:2019cqn} states that given a toric diagram associated to a toric geometry $X$, one can compute the vectors $p_i$, obtained as the differences $v_{i+1}-v_i$ of lattice points on the perimeter of the toric diagram, taken to be the two extrema of each edge\footnote{I.e.\ ignore the vertices along the edge which are not its extrema.}. One should also compute the quantity $g_{ij} = \text{det}(p_i,p_j)$, for each pair of consecutive edges. Then the singular toric threefold at hand is an orbifold of some other singularity via the action of a group of order $n$, with $n=\gcd(g_{12},g_{23},\dots)$. They then further observe that it is possible to construct a dimer model from said toric diagram enjoying a order $n$ discrete symmetry. Since dimer models are nothing but a combinatorial tool to encode both the BPS quiver and the superpotential, one can interpret this discrete symmetry as the quantum symmetry arising after an obifold. Indeed, as reviewed in \cref{app:toric_theory}, it is a standard result in toric geometry that an orbifold of order $n$, that preserves the toricity of the geometry, will produce a new toric diagram with an area $n$ times larger than the starting one.

This result can be used in conjunction with the findings of \cite{Albertini:2020mdx, Morrison:2020ool}. The higher form symmetries are determined by $\tilde n = \gcd(\tilde g_{ij}, ...)$, where now the $\tilde p_i$ are computed from the difference of two consecutive vertices $v_i$, counting each point along one edge, not just the endpoints. By construction $\tilde n | n$, and thus one can unorbifold a normal subgroup of order $\tilde n$ to get an effective orbifold of order $n/\tilde n$, engineering a theory without higher form symmetries. To do so explicitly, one needs to  know the action of the orbifold on the coordinates describing $X$ and repeat the procedure described in the previous section.

Let us show this recipe concretely in action. We consider the Calabi-Yau threefold $X$ to be described by the 2d toric diagram $\{ (0,0),(2,0),(3,4),(2,4) \}$, where we omit the third coordinate of the vectors, since they all lie on the same plane. By applying the above analysis we see that $n=8$ and $\tilde n = 4$. This theory is indeed an orbifold of the conifold, $Y = \{ (0,0),(1,0),(1,1),(0,1) \}$ and $X = Y/\mathbb{Z}_8$. In order to single out the elements of the orbifold group that act with fixed loci we now have to specify the $\mathbb{Z}_8$ action on the coordinates $(x,y,z,w) \in \C[x,y,z,w]$, with $\C[x,y,z,w]/(xy-zw)$ the coordinate ring of the conifold. We take 
\begin{align}
    g: (x \to \omega^2 x \, , y \to \omega^6 x \, , z \to \omega^7 z \, , w \to \omega^7 w )\, ,
\end{align}
with $\omega^8=1$. From this action, it is easy to see that $g^4$ acts trivially on $(x,y)$, and hence we can extract from the character table of $\Z_8$ the irreps on which $g^4$ is trivialized, thus forming the quantum subgroup that encodes the higher form symmetry. Finally, one can ungauge this subgroup and is left with a theory with a trivial Defect Group, which is indeed the 5d SCFT corresponding to the conifold. 

For a generic toric diagram, this procedure is not as straightforward since the equations defining the threefold are in general a non-complete intersection in a suitable affine space. However, starting from a theory corresponding to a generic toric diagram $X$, it is still possible to unorbifold the whole quantum dual to reduce to the theory $Y$, which by definition is a theory with a trivial Defect Group. We can however provide an heuristic argument on how to single out the subgroup acting without fixed points.

Suppose one has the toric diagram for $X = Y / \Gamma$ and $Y$. We know that in $Y$ there are two consecutive vectors $p_i$ such that det$(p_i,p_{i+1})=\pm1$, since by construction the theory is not an orbifold of something else. We can thus use these vectors as the ones spanning the lattice in which we draw the toric diagram of $Y$ and encode the matrix that maps these base vectors into the ones of $X$. In the example above we can take $p_1=(1,0)-(0,0)$ and $p_2=(0,1)-(0,0)$ and thus the orbifold matrix reads
\begin{align}
    \text{M}_{orb} = \begin{pmatrix}
        2 & 1  \\
        0 & 4 \\
    \end{pmatrix} \,.
\end{align}
One can check that the cokernel of map associated to $M_{orb}$ is $\Z_8$, as expected.

The above matrix acts on the $p_i$ of $Y$ from the left: in particular it acts on $p_1$ just by multiplication, thus introducing some extra vertex along the perimeter of the toric diagram of $X$. This is the hallmark of a non-isolated singularity, meaning that the orbifold is acting with some fixed point action. In general, if the image of a $p_i$ is of the from $l \cdot q_i$, for some vector $q_i$ and integer $l$, then there is a $\Z_l$ with fixed loci action of the orbifold. Thus, the un-orbifolded theory will be of the type $X' = Y/\Z_l$ for $\Z_l \subset \Gamma$.
 
Let us conclude by saying that, although in the cases of abelian orbifolds of $\mathbb{C}^3$ the quantum dual is readily given as a cyclic symmetry of the McKay quiver, this is not a general feature. The explicit case by case analysis for an arbitrary toric threefold might depend on the explicit mutation of the quiver, where the quantum symmetry might or might not be explicitly realized. Nonetheless, the above results are general and don't rely on the choice of a mutant, but only of the geometric data.

In what follows we will provide further examples of higher form symmetries arising through orbifolding also in non toric cases. We do not provide a general proof of the orbifold origin of the defect group, despite proving it in the toric case. We are confident that our analysis should generalize to any non-compact CY geometry, leaving the proof to future work.

\subsection{Group extensions and higher groups}

We conclude this section commenting on how \cref{eq:exact_quantum_group} encodes possible higher group structures of the field theory. This connection was already pointed out in \cite{DelZotto:2022fnw, Cvetic:2022imb}, but we try to interpret it in terms of a quantum symmetry of the BPS quiver.

When the flavour symmetry is faithfully realized in the geometry,\footnote{The actual flavor symmetry is in general more complicated since the naive geometric one can undergo enhancement. We will focus only on theory in which the flavor symmetry matches the one that can be read from the toric diagram.} the higher group structure was described in \cite{DelZotto:2022joo, Cvetic:2022imb}. To read the higher group structure from the toric diagram one defines the structure group $\mathcal{E}$ in terms of the boundary geometry and checks if the following exact sequence splits 
\begin{align}
  0 \to  \mathbb{D}^{(1)} \to \mathcal{E} \to \mathcal{F} \to 0 \, ,
\end{align}
where $\mathbb{D}^{(1)}$ is the $1$-form symmetry group of the theory.

Following the above references, the way by which the structure group is computed is completely equivalent to the computation of the quantum symmetry we described in the previous sections. Therefore, we can identify the above sequence encoding higher group structures with the one dual to \cref{eq:exact_quantum_group}. Thus we conclude that the 2-group structure arising in toric 5d SCFTs can be fully read from the lattice of $Y/\Gamma$ and the sublattice of $Y$ as long as there is no symmetry enhancement.

%\section{More involved examples, conifold and toric singularities}

\section{Examples}\label{sec:examples}

In this section we exhibit explicit examples of the orbifolding procedure outlined in the previous sections. We start from a (possibly non-toric) canonical CY$_3$ $X$ and explore the 5d SCFTs engineered by the orbifolds of $X$. We guarantee that the threefold that results from the orbifold is a canonical CY$_3$, imposing that the orbifold action should:
\begin{itemize}
    \item be an isometry of $X$;
    \item preserve the holomorphic 3-form of the original threefold $X$.
\end{itemize}
As we have previously remarked, the BPS quiver of these 5d SCFTs theories also encodes their Defect Group. Moreover, as a welcome bonus, we extract salient features such as the gauge and flavor rank of these 5d SCFTs without the need of an explicit resolution of the singularity. As reviewed in section \ref{sec: 5d SCFTs}, in the rest of the section we refer to $r$ as the gauge rank of the theory and to $f$ as its flavour rank.

\subsection{Toric examples: orbifolds of the conifold}
We gently start by revisiting classic examples of abelian orbifolds of the conifold \cite{Oh:1999sk}, which is the prototypical toric isolated singularity. We are particularly interested in the algebraic presentation of its abelian orbifolds, as well as in their BPS quiver. This data encodes the 5d SCFT realized via M-theory geometric on the orbifold, dictating its gauge and flavor symmetry, along with its 1-form symmetry.

Consider the usual embedding of the conifold as a hypersurface in the affine space $\mathbb{C}^4$:
\begin{equation}\label{eq:conifold CY3}
    x y = z w \quad \subset \mathbb{C}^4.
\end{equation}
It is well-known that the conifold admits only a small crepant resolution, inflating a compact curve isomorphic to $\mathbb{P}^1$. Hence:
\begin{equation}
    r = 0, \quad f = 1.
\end{equation}
The 5d SCFT engineered by the conifold is therefore a rank-0 theory describing a free hypermultiplet\footnote{See \cite{Acharya:2024bnt} for recent progress highlighting subtleties in this description. These will not affect our conclusions.}.\\
Its BPS quiver is as in Figure \ref{fig:conifold quiver}. Here and onwards we label the quiver nodes according to the notation of appendix \ref{app:skew_group} and \cite{Aspinwall:2010mw}\footnote{$R$  is the ring defining the threefold, and $M$ is the module corresponding to a matrix factorization of the threefold. When introducing an orbifold action, we will add labels $R_i$ and $M_i$ to distinguish the multiple copies of the original nodes. In other instances, we simply label the nodes with numbers.}.
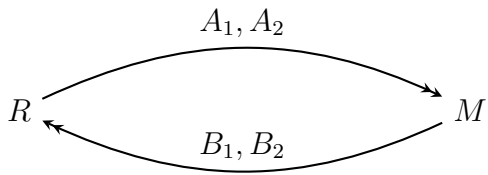
\begin{figure}[H]
\begin{center}
\begin{tikzpicture}[>=stealth, thick]
% Nodes
\node (R) at (0,0) {$R$};
\node (M) at (6,0) {$M$};
% R → M arrows (upper and lower)
%\draw[->] (R) to[out=50, in=130] node[above]{$A_1$} (M);
%\draw[->] (M) to[out=-130, in=-50] node[above]{$B_1$} (R);
\draw[->>] (R) to[out=25, in=155] node[above]{$A_1,A_2$} (M);
\draw[->>] (M) to[out=-155, in=-25] node[above]{$B_1,B_2$} (R);
\end{tikzpicture}\\
\end{center}
\caption{BPS quiver for the conifold geometry.}
\label{fig:conifold quiver}
\end{figure}

%\begin{equation}\label{conifold quiver}
%\begin{tikzpicture}[>=stealth, thick]
% Nodes
%\node (R) at (0,0) {$R$};
%\node (M) at (6,0) {$M$};
% R → M arrows (upper and lower)
%\draw[->] (R) .. controls (3,2) .. node[above=2pt] {$A_1$} (M);
%\draw[->] (R) .. controls (3,0.5) .. node[above=2pt] {$A_2$} (M);
% M → R arrows (upper and lower, more curved so they don't overlap)
%\draw[->] (M) .. controls (3,-0.5)  .. node[below=2pt] {$B_1$} (R);
%\draw[->] (M) .. controls (3,-2)  .. node[below=2pt] {$B_2$} (R);
%\end{tikzpicture}\\
%\end{equation}
The superpotential reads:
\begin{equation}
    W = \text{Tr}(\epsilon_{il}\epsilon_{jk}A_iB_j A_l B_k),
\end{equation}
and the relations between arrows and coordinates in the ring $\mathbb{C}[x,y,z,w]$ are:
\begin{equation}
    x = A_1B_1 \, , \quad y = A_2B_2 \, , \quad z = A_1B_2 \, , \quad w = A_2B_1 \, . 
\end{equation}

We implement a possible choice of abelian orbifold action as:
\begin{equation}\label{eq:orbifold conifold}
    (x,y,z,w) \rightarrow (\lambda_1 x,\lambda_1^{-1}y,\lambda_2 z,\lambda_2^{-1} w) \, , \quad \lambda_1 \in \mathbb{Z}_{n_1} \, , \, \,  \lambda_2 \in \mathbb{Z}_{n_2}.
\end{equation}
It is immediate to check that \cref{eq:conifold CY3} is preserved under this action, along with the holomorphic 3-form $\Omega_{3,0} = \text{Res}_{\{xy-zw=0\}}\frac{dx\wedge dy\wedge dz\wedge dw}{xy-zw}$. Hence, the resulting orbifold satisfies the Calabi-Yau condition.

Let us compute the CY$_3$ that is produced by the orbifold action. A minimal set of generators of invariants under the action in \cref{eq:orbifold conifold} are:
\begin{equation}
\begin{split}
  &  a_1 = x^{n_1},\quad a_2 = y^{n_1}, \quad a_3 = xy,\\
  &  a_4 = z^{n_2}, \quad a_5 = w^{n_2}, \quad a_6 = zw.\\
    \end{split}
\end{equation}
The relations between the invariants, supplemented by the conifold equation \eqref{eq:conifold CY3}, read:
\begin{equation}\label{eq:conifold relations}
    \begin{cases}
        a_1a_2=a_3^{n_1}\\
        a_4a_5=a_6^{n_2}\\
        a_3 = a_6\\
    \end{cases}.
\end{equation}
Notice that the conifold equation, rewritten in terms of the invariant coordinates, lies at the third row of \cref{eq:conifold relations}. Discarding spurious equations, the Calabi-Yau threefold resulting from the orbifold is:
\begin{equation}\label{eq:conifold orbifold CY3}
    \begin{cases}
        a_1a_2 = a_3^{n_1}\\
        a_4a_5 = a_3^{n_2}\\
    \end{cases}.
\end{equation}
This is a toric CY$_3$ encoded by the toric diagram:
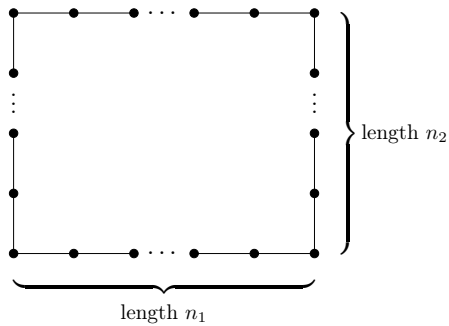
\begin{figure}[H]
\centering
    \scalebox{0.8}{
    \begin{tikzpicture}
        \draw (0,2)--(0,0)--(2,0);
        \draw (3,0)--(5,0)--(5,2);
        \draw (5,3)--(5,4)--(3,4);
        \draw (2,4)--(0,4)--(0,3);
        \filldraw (0,0) circle (2pt);
        \filldraw (1,0) circle (2pt);
        \filldraw (2,0) circle (2pt);
        \filldraw (3,0) circle (2pt);
        \filldraw (4,0) circle (2pt);
        \filldraw (5,0) circle (2pt);
        \filldraw (5,1) circle (2pt);
        \filldraw (5,2) circle (2pt);
        \filldraw (5,3) circle (2pt);
        \filldraw (5,4) circle (2pt);
        \filldraw (0,4) circle (2pt);
        \filldraw (1,4) circle (2pt);
        \filldraw (2,4) circle (2pt);
        \filldraw (3,4) circle (2pt);
        \filldraw (4,4) circle (2pt);
        \filldraw (0,1) circle (2pt);
        \filldraw (0,2) circle (2pt);
        \filldraw (0,3) circle (2pt);
        \node at (0,2.6) {$\vdots$};
        \node at (5,2.6) {$\vdots$};
        \node at (2.5,0) {$\cdots$};
        \node at (2.5,4) {$\cdots$};
      \node at (2.5,-0.7)  {$\underbrace{\hspace{5cm}}_{}$};
       \node at (5.7,2) [rotate = 90]       {$\underbrace{\hspace{4cm}}_{}$};
       \node at (2.5,-1) {\footnotesize length $n_1$};
       \node at (6.5,2) {\footnotesize length $n_2$};
    \end{tikzpicture}}
    \caption{Toric diagram for the CY$_3$ in \cref{eq:conifold orbifold CY3}.}
    \label{fig:conifoldorbifoldCY3}
    \end{figure}

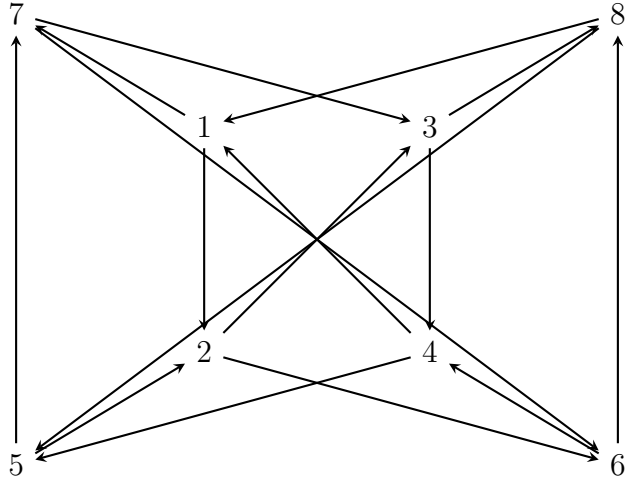
\begin{figure}[H]
\begin{center}
\begin{tikzpicture}[>=stealth, thick]
% Nodes
\node (1) at (-1.5,1.5) {$1$};
\node (3) at (1.5,1.5) {$3$};
\node (2) at (-1.5,-1.5) {$2$};
\node (4) at (1.5,-1.5) {$4$};
\node (7) at (-4,3) {$7$};
\node (8) at (4,3) {$8$};
\node (5) at (-4,-3) {$5$};
\node (6) at (4,-3) {$6$};
% Loops
\draw[->] (1) to (2);
\draw[->] (1) to (7);
\draw[->] (2) to (3);
\draw[->] (2) to (6);
\draw[->] (3) to (4);
\draw[->] (3) to (8);
\draw[->] (4) to (1);
\draw[->] (4) to (5);
\draw[->] (5) to (2);
\draw[->] (5) to (7);
\draw[->] (6) to (4);
\draw[->] (6) to (8);
\draw[->] (7) to (3);
\draw[->] (7) to (6);
\draw[->] (8) to (1);
\draw[->] (8) to (5);
\end{tikzpicture}
\end{center}
\caption{BPS quiver for the geometry in \cref{eq:conifold orbifold CY3}, with $n_1 = n_2 = 2$.}
\label{fig:box quiver}
\end{figure}

The BPS quiver for the 5d SCFT engineered by \cref{eq:conifold orbifold CY3} can be straightforwardly computed thanks to the techniques reviewed in \cref{app:skew_group}, given the explicit action of the orbifold in \cref{eq:orbifold conifold} and its translation in terms of an action on the arrows of Figure \ref{fig:conifold quiver}. Finally, the BPS quiver is displayed in Figure \ref{fig:box quiver}.

It can be easily checked that it agrees with the result obtained from the brane-tiling of the toric CY$_3$ depicted in \Cref{fig:conifoldorbifoldCY3}.

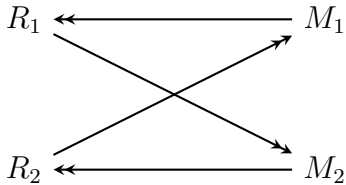
\begin{figure}[H]
\begin{center}
\begin{tikzpicture}[>=stealth, thick]
% Nodes
\node (1) at (-2,1) {$R_1$};
\node (3) at (2,1) {$M_1$};
\node (2) at (-2,-1) {$R_2$};
\node (4) at (2,-1) {$M_2$};
% Loops
\draw[->>] (3) to (1);
\draw[->>] (4) to (2);
\draw[->>] (2) to (3);
\draw[->>] (1) to (4);
\end{tikzpicture}
\end{center}
\caption{BPS quiver for the local $\mathbb{F}_0$ geometry.}
\label{fig:F0 quiver}
\end{figure}

On top of the above actions, that do not generate higher form symmetries, one can consider the following orbifold with diagonal action:
\begin{equation}\label{eq:orbifold conifold 2}
    (x,y,z,w) \rightarrow (\lambda x,\lambda^{-1}y,\lambda z,\lambda^{-1} w), \quad \lambda \in \mathbb{Z}_{n}.
\end{equation}
If we take for example $\mathbb{Z}_{2}$ as above, we have that the resulting theory is engineered by the local $\mathbb{F}_0$ threefold.  The invariant coordinates read:
\begin{equation}\label{eq:invariants F0}
\begin{array}{cccc}
        a_1 = x^2, & a_2 = y^2, & a_3 = z^2, & a_4 = w^2,\\
    a_5 = xy, & a_6 = xz, & a_7 = xw, & a_8 = yz,\\
    a_9 = yw, & a_{10} = zw. & &\\
   \end{array}
\end{equation}
Computing the independent relations between the invariants in \cref{eq:invariants F0}, we can define the orbifolded threefold, i.e.\ local $\mathbb{F}_0$, as a non-complete intersection in $\mathbb{C}^{10}$:
\begin{equation}\label{eq:local F0}
\begin{cases}
\begin{array}{cccc}
a_1a_2 = a_5^2, \quad & a_3a_4 = a_{10}^2,\quad & a_5 a_8 = a_2 a_6,\quad & a_8 a_{10} = a_3 a_9,\\
a_1a_3 = a_6^2, \quad & a_5 a_{10} = a_6 a_9, \quad & a_5a_9 = a_2a_7, \quad & a_7 a_9 = a_4 a_5,\\
a_1a_4 = a_7^2, \quad & a_5a_{10} = a_7a_8, \quad & a_6a_7 = a_1a_{10}, \quad & a_7a_{10} = a_4a_6,\\
a_2a_3 = a_8^2, \quad & a_5a_6 = a_1a_8, \quad & a_6a_8 = a_3a_5, \quad & a_8 a_9 = a_2a_{10},\\
a_2a_4 = a_9^2 ,\quad & a_5a_7 = a_1a_9, \quad & a_6a_{10} = a_3a_7, \quad & a_9a_{10} = a_4a_8,\\
\boxed{a_5 = a_{10} } .\quad &&&\\
\end{array}
\end{cases}
\end{equation}
Notice that the boxed relation in \cref{eq:local F0} is the conifold defining equation, rewritten in terms of the invariants under the orbifold action: it supplies a linear constraint on the other quadratic relations. This algebraic analysis agrees with the result of the previous section: indeed the action of the orbifold is free of fixed loci, and thus a non trivial $\mathbb{Z}_2$ Defect Group is generated. The BPS quiver is as in Figure \ref{fig:F0 quiver}.

More in general, all $Y^{p,0}$ theories can be obtained through this kind of quotient. The set of polynomials defining the singular threefold is increasingly hard to compute directly as the $\mathbb{Z}_{n}$ invariant subring, but it can be read from the toric diagram instead. For a systematic approach to the computation of the invariant subring, see \cite{Pinansky:2005ex}.

Consistently with the results of the previous sections, we notice that the theories with a non trivial defect group are precisely the ones obtained by orbifolding the Conifold without fixed loci, i.e.\ the $Y^{p,0}$ geometries.

\subsection{Finite Abelian orbifolds of cDV singularities}
In this Section we move on to \textit{non-toric} threefolds, and explore examples of abelian orbifolds of compound Du Val (cDV) singularities. These theories were recently constructed in \cite{Moleti}, to which we refer the reader for in-depth geometric details. Here we are chiefly concerned with testing our conjecture and computing the defect group of the corresponding 5d SCFTs.\\

The BPS quiver of cDV singularities is either known or can be straightforwardly computed via the matrix factorization machinery developed in \cite{Aspinwall:2010mw}. cDV singularities are defined as follows:
\begin{equation}
    P_{\mathfrak{g}}(x,y,z) + wf(x,y,z,w) = 0 \quad \subset\mathbb{C}^4, 
\end{equation}
with $P_{\mathfrak{g}}(x,y,z)$ a polynomial defining a Du Val singularity of type $\mathfrak{g}$,\footnote{Recall that Du Val/Kleinian/ADE singularities, or rational double points, which are the only 2-complex dimensional canonical singularities, can be defined as follows:
\begin{equation}
    \begin{split}
        & A_n: \quad xy-z^{n+1}=0,\\
        & D_n: \quad x^2+zy^2+z^{n-1}=0,\\
        &E_6: \quad x^2+y^3+z^4 =  0,\\
        &E_7: \quad x^2+y^3+yz^3 =  0,\\
        &E_8: \quad x^2+y^3+z^5 =  0.\\
    \end{split}
\end{equation}} and $f(x,y,z,w)$ a polynomial function. As is well-known thanks to a theorem by Reid \cite{reidyoungpersonsing}, cDV singularities admit at most small resolutions. Thanks to the geometric engineering dictionary reviewed in \cref{sec: 5d SCFTs}, this implies that the corresponding 5d SCFTs are always rank 0. We will see that this \textit{does not} necessarily hold for orbifolds of cDV singularities, as these are often not cDV singularities themselves.

\subsubsection{c$\boldsymbol{A}$ singular threefolds}
In this section we focus on a class of c$A$ singular threefolds with an isolated singularity, admitting a small crepant resolution with a collection of $\mathbb{P}^1$'s as exceptional locus. They can be presented as a hypersurface in $\mathbb{C}^4$:
\begin{equation}\label{eq:cAn hypersurface}
X: \quad   xy = z^{2n}+w^{2k} \quad \subset \mathbb{C}^4, 
\end{equation}
where we choose $n\geq k$, with no loss of generality. As such, these threefolds engineer rank-0 5d SCFTs, with flavor group of rank $gcd(2n,2k)-1$.
This is exactly $\text{dim}H_2(\tilde{X},\mathbb{R})$, where $\tilde{X}$ is a suitably chosen crepant resolution of $X$. In general, when $k \nmid n$, there are leftover terminal singularities, and in such cases $\tilde{X}$ is only a partial resolution.

\begin{figure}[H]
\begin{center}
\begin{tikzpicture}[>=stealth, thick]
% Nodes
\node (1) at (0,0) {$1$};
\node (2) at (2,-2) {$2$};
\node (3) at (0,-4) {$3$};
\node (4) at (-2,-2) {$4$};
% Loops
\draw[->] (1) edge[->, min distance=12mm, in=60, out=120] node[above]{$\Phi_1$} (1);
\draw[->] (2) edge[->, min distance=12mm, in=30, out=-30] node[right]{$\Phi_2$} (2);
\draw[->] (3) edge[->, min distance=12mm, in=-60, out=-120] node[below]{$\Phi_3$} (3);
\draw[->] (4) edge[->, min distance=12mm, in=150, out=-150] node[left]{$\Phi_4$} (4);
%%%%%
\draw[->, transform canvas={xshift=3pt,yshift=3pt}] (2)--(1)   node[midway,xshift=12pt] {$B_{21}$};
\draw[->, transform canvas={yshift=-3pt}] (1)--(2) node[midway,xshift=-10pt,yshift=-3pt] {$A_{12}$};
\draw[->, transform canvas={xshift=-3pt,yshift=3pt}](1)--(4)node[midway,xshift=-14pt] {$B_{14}$};
\draw[->, transform canvas={yshift=-3pt}] (4)--(1)node[midway,xshift=10pt,yshift=-3pt] {$A_{41}$};
\draw[->, transform canvas={yshift=3pt}] (2)--(3)node[midway,xshift=-8pt,yshift=10pt] {$A_{32}$};
\draw[->, transform canvas={xshift=3pt,yshift=-3pt}] (3)--(2)node[midway,yshift=-13pt] {$B_{23}$};
\draw[->, transform canvas={yshift=3pt}] (3)--(4)node[midway,xshift=3pt,yshift=10pt] {$A_{34}$};
\draw[->, transform canvas={xshift=-3pt,yshift=-3pt}] (4)--(3)node[midway,yshift=-10pt] {$B_{43}$};
\end{tikzpicture}
\end{center}
    \caption{BPS quiver for $\mathbb{C}^2/\mathbb{Z}_4\times \mathbb{C}$.}
    \label{fig:placeholderdp}
\end{figure}
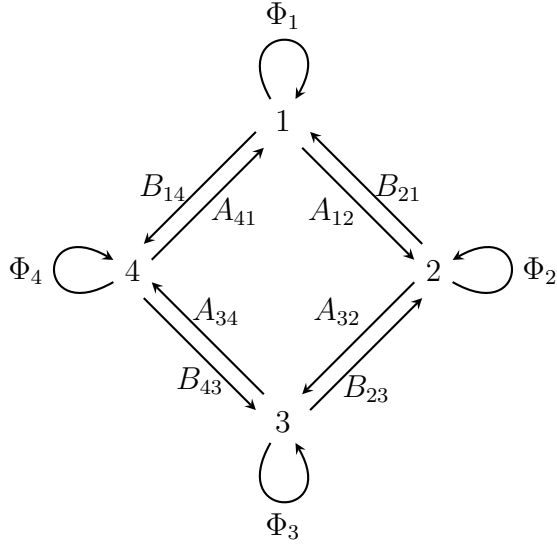

The BPS quiver for these theories was computed in \cite{Aspinwall:2010mw}, and it turns out to be the same as the one for the $\C^2/\Z_{2\gcd(k,n)}\times \C$ with a deformed superpotential. For example, in the case of $n=k=2$ we have the quiver in \Cref{fig:placeholderdp}, with the superpotential:
\begin{equation}\label{eq:supo_cDVA}
    W = \sum_{i=1}^4 \Phi_i(A_{i,i+1}B_{i+1,i} - B_{i,i-1}A_{i-1,i}) - 2i \Phi_1^{2} + (i+1) \Phi_2^{2} - 2 \Phi_3^{2} + (i+1) \Phi_4^{2} \, .
\end{equation}
As it can be easily checked, the moduli space of the associated field theory is indeed described by \cref{eq:cAn hypersurface}.

The most general abelian orbifold action that preserves the nowhere-vanishing 3-form, as well as the threefold defined by \cref{eq:cAn hypersurface}, is:
\begin{equation}\label{eq:orbifold CA}
     (x,y,z,w) \rightarrow (\lambda\mu x,\lambda^{-1}\mu y,\mu z,\mu w), \quad \lambda \in \mathbb{Z}_m, \mu \in \mathbb{Z}_2.
\end{equation}
Focusing on the $m=2$ case, the invariant coordinates are exactly the same as the ones listed in \cref{eq:invariants F0} and the orbifolded threefold is given by the non-complete intersection:
\begin{equation}\label{eq:local cAn}
\begin{cases}
\begin{array}{cccc}
a_1a_2 = a_5^2, \quad & a_3a_4 = a_{10}^2,\quad & a_5 a_8 = a_2 a_6,\quad & a_8 a_{10} = a_3 a_9,\\
a_1a_3 = a_6^2, \quad & a_5 a_{10} = a_6 a_9, \quad & a_5a_9 = a_2a_7, \quad & a_7 a_9 = a_4 a_5,\\
a_1a_4 = a_7^2, \quad & a_5a_{10} = a_7a_8, \quad & a_6a_7 = a_1a_{10}, \quad & a_7a_{10} = a_4a_6,\\
a_2a_3 = a_8^2, \quad & a_5a_6 = a_1a_8, \quad & a_6a_8 = a_3a_5, \quad & a_8 a_9 = a_2a_{10},\\
a_2a_4 = a_9^2 ,\quad & a_5a_7 = a_1a_9, \quad & a_6a_{10} = a_3a_7, \quad & a_9a_{10} = a_4a_8,\\
\boxed{a_5 = a_3^n+a_4^k } .\quad &&&\\
\end{array}
\end{cases}
\end{equation}
The boxed relation in \cref{eq:local cAn} is the hypersurface in \cref{eq:cAn hypersurface}, rewritten in terms of orbifold-invariant coordinates, giving a linear constraint on the other relations. Let us remark a key fact: the orbifolded threefold in \cref{eq:local cAn} is almost completely identical to the presentation of the local $\mathbb{F}_0$ geometry in \cref{eq:local F0}, that we have obtained as a $\mathbb{Z}_2$ quotient of the conifold.
The crucial difference is that the boxed relation in \cref{eq:local cAn} is a sum, and as such it implies that the corresponding threefold is non-toric. We then see that orbifolding this class of c$A$ singularities naturally produces Calabi-Yau threefolds\footnote{The threefolds are guaranteed to satisfy the Calabi-Yau condition, since the orbifold preserves the holomorphic 3-form, as well as the defining equation of the non-orbifolded threefold.} which are non-toric non-complete intersections. This class of threefolds has been rarely investigated in the framework of Geometric Engineering, even though it most likely comprises a large swath of canonical threefolds, that remain to this day mostly unexplored.

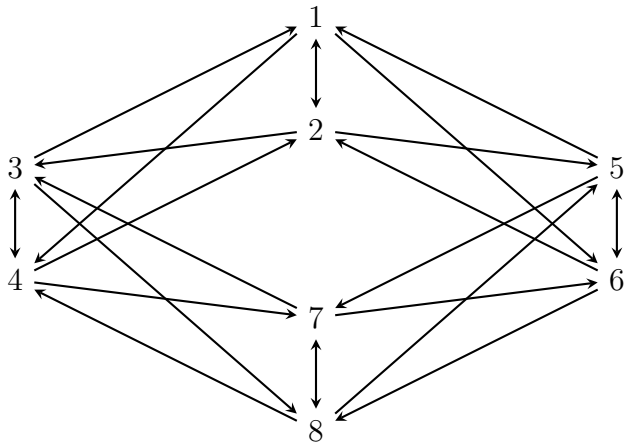
\begin{figure}[H]
\begin{center}
\scalemath{1}{
\begin{tikzpicture}[>=stealth, thick]
% Nodes
\node (R_1) at (0,0) {$1$};
\node (M_1) at (-4,-2) {$3$};
\node (M_2) at (4,-2) {$5$};
\node (M_3) at (0,-4) {$7$};
%%%%
\node (R_2) at (0,-1.5) {$2$};
\node (M_4) at (-4,-3.5) {$4$};
\node (M_5) at (4,-3.5) {$6$};
\node (M_6) at (0,-5.5) {$8$};
%%%%
\draw[->] (M_1) to (R_1);
\draw[->] (M_2) to (R_1);
\draw[->] (M_5) to (M_6);
\draw[->] (M_6) to (M_4);
\draw[->] (R_2) to (M_1);
\draw[->] (R_2) to (M_2);
\draw[->] (M_5) to (R_2);
\draw[->] (M_4) to (R_2);
\draw[->] (M_3) to (M_1);
\draw[->] (M_1) to (M_6);
\draw[->] (M_6) to (M_2);
\draw[->] (M_2) to (M_3);
\draw[->] (R_1) to (M_5);
\draw[->] (R_1) to (M_4);
\draw[->] (M_3) to (M_5);
\draw[->] (M_4) to (M_3);
%%%
\draw[<->] (M_1) to (M_4);
\draw[<->] (M_2) to (M_5);
\draw[<->] (R_1) to (R_2);
\draw[<->] (M_3) to (M_6);
%%%%
%\draw[->] (R_2) to[out=50, in=-145] node[above, xshift=22pt, yshift=20pt]{$C_1$} (M_1);
\end{tikzpicture}
}
\end{center}
\caption{BPS quiver for the geometry in \cref{eq:local cAn}, with $n=2$ and $k=2$.}
\label{fig: cA quiver}
\end{figure}

We can now come to the analysis of the BPS quivers and 1-form symmetries of the orbifolded threefolds.
 The BPS quiver for the geometry described by \cref{eq:local cAn} when $p = \gcd(n,k) = \min(n,k)$ is given by the quiver for $\C^3/\Z_{4p}$ with action $(1,2p-1,2p)$, with the same superpotential of the orbifold plus a $m = \max(n,k)$ power term for invariant fields obtained by the 2-cycles in the quiver. For example, for $(n,k)=(2,2)$ we end up with the quiver for $\C^3/\Z_8$, with action $(1,3,4)$, whose quiver is given in \Cref{fig: cA quiver}.
Then, the corresponding superpotential is (denoting by $\Phi_{ij}$ the arrow from node $i$ to $j$):
\begin{align}
    W_{tot} = W_{orb} - 2i (\Phi_{12} \Phi_{21})^{2} + (i+1) (\Phi_{34} \Phi_{43})^{2} - 2 (\Phi_{56} \Phi_{65})^{2} + (i+1) (\Phi_{78} \Phi_{87})^{2} \, , 
\end{align}
as can be checked perfoming the orbifold projection of the potential in \cref{eq:supo_cDVA}.

From the 5d point of view, this BPS quiver describes a 5d SCFT with:
\begin{equation}
    r = 2 \, , \quad\quad f=3 \, .
\end{equation}
As can be easily computed, in the theory that produces the BPS quiver in Figure \ref{fig: cA quiver} there is a non trivial $\Z_2$ defect group, in agreement with the general orbifold analysis. More in general, one has that the resulting theory shares the same gauge and flavour rank with the undeformed $\C^3/\Z_{4p}$ one, having thus
\begin{equation}
    r = p \, , \quad\quad f=2p-1 \, .
\end{equation}

\subsubsection{Laufer's singularity}
\begin{figure}[H]
\begin{center}
\scalemath{1}{
\begin{tikzpicture}[>=stealth, thick]
% Nodes
\node (R) at (0,0) {$R$};
\node (M) at (6,0) {$M$};
% Loops
\draw[->] (R) .. controls +(150:1.8) and +(210:1.8) .. node[left] {$D$} (R);
\draw[->] (M) .. controls +(30:4.5) and +(-30:4.5) .. node[right] {$C_2$} (M);    % huge loop
\draw[->] (M) .. controls +(30:2.3) and +(-30:2.3) .. node[right, xshift=2pt] {$C_1$} (M); % large loop
% R → M arrows (upper and lower)
\draw[->] (R) to[out=25, in=155] node[above]{$A$} (M);
\draw[->] (M) to[out=-155, in=-25] node[above]{$B$} (R);
\end{tikzpicture}
}
\end{center}
\caption{BPS quiver for the geometry in \cref{eq:Laufer sing}.}
\label{fig:laufer quiver}
\end{figure}
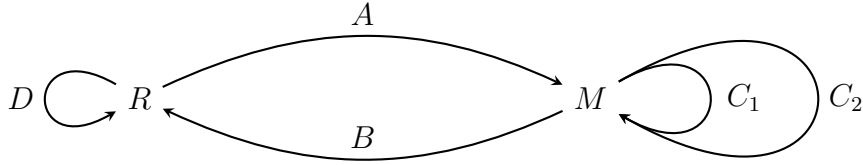

In this section we delve into cDV singularites of type c$D$, focusing on a class of threefolds with isolated singularities particularly amenable to our purposes, known as Laufer's singularities.\footnote{Notice that this technology can be neatly generalized to all cDV singularities that possess a NCCR \cite{Aspinwall:2010mw}.}  This threefold is defined by the equation:
\begin{equation}\label{eq:Laufer sing}
    x^2 + zy^2 + w^3+w z^{2n+1} = 0 \quad \subset\mathbb{C}^4 \, ,
\end{equation}
which has only an isolated singularity at the origin. Its BPS quiver is shown in Figure \ref{fig:laufer quiver}.
The superpotential is:
\begin{equation}
    W = ADB+ABC_2^2+C_1^2C_2+\frac{D^{n+1}}{n+1}+\frac{(-1)^n}{2n+2}C_2^{2n+2} \, ,
\end{equation}
where the relation between coordinates and arrows is:
\begin{equation}
 \begin{split}
     & z = -C_2^2 \, ,\\
     & x = ABC_2C_1+C_2C_1AB+(-1)^mC_2^{2n+1}C_1 \, ,\\
     & w = -C_1^2 \, ,\\
     & y = -ABC_1-C_1AB-(-1)^mC_1C_2^{2n} \, . \\
 \end{split}   
\end{equation}
As an application of our techniques, let us consider the following orbifold action, that preserves the holomorphic 3-form as well as the zero locus of \cref{eq:Laufer sing}:
\begin{equation}\label{laufer Z2}
    (x,y,z,w) \rightarrow (\lambda x,\lambda y, z, w), \quad \lambda \in \mathbb{Z}_2.
\end{equation}

\begin{figure}[H]
\begin{center}
\begin{tikzpicture}[>=stealth, thick]
% Nodes
\node (R_1) at (0,0) {$R_1$};
\node (M_1) at (4,0) {$M_1$};
\node (R_2) at (-4,0) {$R_2$};
\node (M_2) at (8,0) {$M_2$};
% Loops
\draw[->] (R_1) edge[->, min distance=12mm, in=45, out=135] node[above]{$y_1$} (R_1);
\draw[->] (M_1) edge[->, min distance=12mm, in=45, out=135] node[above]{$y_2$} (M_1);
% R → M arrows (upper and lower)
\draw[->] (R_1) to[out=10, in=170] node[above]{$A_2$} (M_1);
\draw[->] (M_1) to[out=-170, in=-10] node[below]{$B_2$} (R_1);
\draw[->] (R_2) to[out=10, in=170] node[above]{$A_1$} (R_1);
\draw[->] (R_1) to[out=-170, in=-10] node[below]{$B_1$} (R_2);
\draw[->] (M_1) to[out=10, in=170] node[above]{$A_3$} (M_2);
\draw[->] (M_2) to[out=-170, in=-10] node[below]{$B_3$} (M_1);
\end{tikzpicture}
\end{center} 
\caption{BPS quiver for the geometry in \cref{eq:laufer orbifold}.}
\label{fig:laufer quiver 2}
\end{figure}
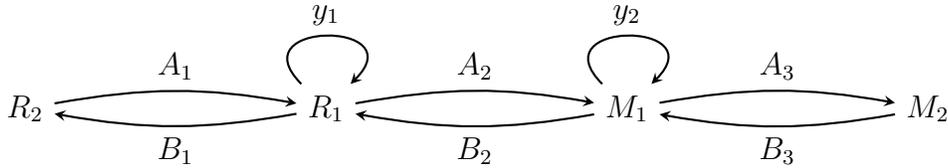

The orbifolded threefold can be easily obtained computing the invariant coordinates:
\begin{equation}
    a_1 = x^2 \, , \quad a_2 = y^2 \, , \quad a_3 = xy \, , \quad a_4 = z \, ,\quad a_5 = w \, ,
\end{equation}
and the corresponding relation:
\begin{equation}\label{eq:laufer orbifold}
    a_3^2+a_4a_2^2 + a_2a_5^3 +a_2 a_5a_4^3 = 0 \quad \subset \mathbb{C}^4.
\end{equation}
Notice that the orbifolded threefold in \cref{eq:laufer orbifold} is a cDV singularity in disguise: this can be swiftly seen by the change of variables $a_2 \rightarrow a_2 + a_4$. This fact is confirmed by the analysis of its BPS quiver. The end result is the quiver in Figure \ref{fig:laufer quiver 2}.
From the Dirac pairing we indeed get the following ranks for the gauge and flavour group respectively:
\begin{equation}
    r = 0 \, , \quad f = 3 \, .
\end{equation}
Hence the 5d SCFT engineered by \cref{eq:laufer orbifold} is rank 0, as expected from the fact that the threefold is cDV. The quiver in Figure \ref{fig:laufer quiver 2} shows that defect group for this orbifold is trivial.\\

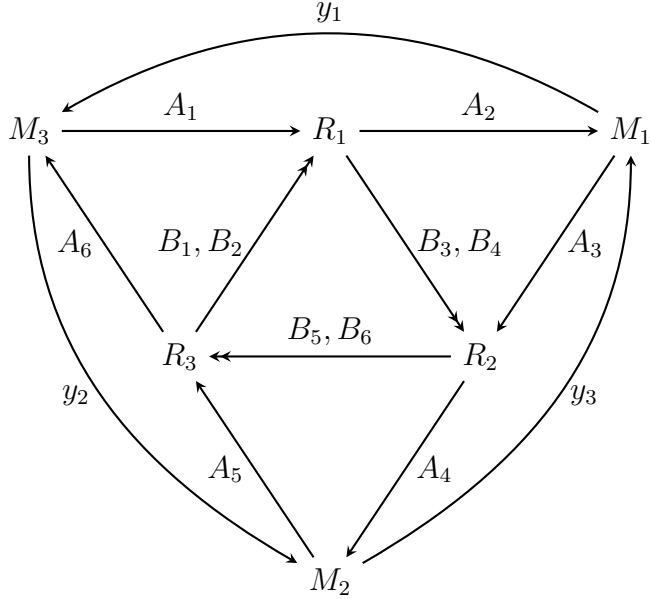
\begin{figure}[H]
\begin{center}
\scalemath{1}{
\begin{tikzpicture}[>=stealth, thick]
% Nodes
\node (R_1) at (0,0) {$R_1$};
\node (M_1) at (-4,0) {$M_3$};
\node (M_2) at (4,0) {$M_1$};
\node (M_3) at (-2,-3) {$R_3$};
\node (M_4) at (2,-3) {$R_2$};
\node (M_5) at (0,-6) {$M_2$};
%%%%
\draw[->] (M_1) to node[above]{$A_1$} (R_1);
\draw[->] (R_1) to node[above]{$A_2$} (M_2);
\draw[->] (M_2) to node[right]{$A_3$} (M_4);
\draw[->] (M_4) to node[right]{$A_4$} (M_5);
\draw[->] (M_5) to node[left]{$A_5$} (M_3);
\draw[->] (M_3) to node[left]{$A_6$} (M_1);
\draw[->>] (M_3) to node[left]{$B_1,B_2$} (R_1);
\draw[->>] (R_1) to node[right]{$B_3,B_4$} (M_4);
\draw[->>] (M_4) to node[above]{$B_5,B_6$} (M_3);

\draw[<-] (M_1) to[out=30, in=150] node[above]{$y_1$} (M_2);
\draw[<-] (M_2) to[out=-90, in=30] node[right]{$y_3$} (M_5);
\draw[->] (M_1) to[out=-90, in=150] node[left]{$y_2$} (M_5);
%\draw[->] (R_2) to[out=50, in=-145] node[above, xshift=22pt, yshift=20pt]{$C_1$} (M_1);
\end{tikzpicture}
}
\end{center}
\caption{BPS quiver for the geometry in \cref{eq:laufer Z3 orbifold}.}
\label{fig:lauferZ3 quiver}
\end{figure}

Consider now a more interesting orbifold action available for Laufer's singularities when $n = 3k-1$,
\begin{equation}\label{eq:laufer Z3}
    (x,y,z,w) \rightarrow (x,\lambda y,\lambda z,\lambda w), \quad \lambda \in \mathbb{Z}_3.
\end{equation}
The coordinates which are invariant under the orbifold action are:
\begin{equation}
\begin{array}{cccc}
    a_1 = x, \quad &a_2 = y^3, \quad& a_3 = w^3, \quad& a_4 = z^3,\\
a_5 = yz^2, \quad &a_6 = y^2z, \quad &a_7 = y w^2, \quad &a_8 = y^2w,\\
 a_9 = z^2w, \quad &a_{10} = zw^2, \quad& a_{11} = y z w.\\
 \end{array}
\end{equation}
The relations between the invariant coordinates are exactly those of the quotient $\mathbb{C}[y,z,w]/\mathbb{Z}_3$, with action $(1,1,1)$, intersected with the orbifold-invariant version of Laufer's singularity (highlighted as the relation in the box of \cref{eq:laufer Z3 orbifold}), which supplies a non-toric linear constraint. The resulting threefold is a non-complete intersection given by 28 equations in $\mathbb{C}^{11}$:
\begin{equation}\label{eq:laufer Z3 orbifold}
    \begin{cases}
        \begin{array}{ccc}
         a_{8} a_{9} - a_{11} a_{11} = 0, \quad &
a_{7} a_{11} - a_{8} a_{10} = 0, \quad &
a_{7} a_{9} - a_{10} a_{11} = 0,\\
a_{6} a_{10} - a_{11} a_{11} = 0, \quad &
a_{6} a_{7} - a_{8} a_{11} = 0, \quad &
a_{5} a_{11} - a_{6} a_{9} = 0,\\
a_{5} a_{10} - a_{9} a_{11} = 0, \quad &
a_{5} a_{8} - a_{6} a_{11} = 0, \quad &
a_{5} a_{7} - a_{11} a_{11} = 0,\\
a_{4} a_{11} - a_{5} a_{9} = 0, \quad &
a_{4} a_{10} - a_{9} a_{9} = 0, \quad &
a_{4} a_{8} - a_{6} a_{9} = 0,\\
a_{4} a_{7} - a_{9} a_{11} = 0, \quad &
a_{4} a_{6} - a_{5} a_{5} = 0, \quad &
a_{3} a_{11} - a_{7} a_{10} = 0,\\
a_{3} a_{9} - a_{10} a_{10} = 0, \quad &
a_{3} a_{8} - a_{7} a_{7} = 0, \quad &
a_{3} a_{6} - a_{8} a_{10} = 0,\\
a_{3} a_{5} - a_{10} a_{11} = 0, \quad &
a_{3} a_{4} - a_{9} a_{10} = 0, \quad &
a_{2} a_{11} - a_{6} a_{8} = 0,\\
a_{2} a_{10} - a_{8} a_{11} = 0, \quad &
a_{2} a_{9} - a_{6} a_{11} = 0, \quad &
a_{2} a_{7} - a_{8} a_{8} = 0,\\
a_{2} a_{5} - a_{6} a_{6} = 0, \quad &
a_{2} a_{4} - a_{5} a_{6} = 0, \quad &
a_{2} a_{3} - a_{7} a_{8} = 0,\\
       \boxed {a_1^2 + a_6 + a_4 + a_9a_3^{2k-1}  = 0}.  & & \\
        \end{array}
    \end{cases}
\end{equation}
The BPS quiver corresponding to the threefold can be neatly computed as in Figure \ref{fig:lauferZ3 quiver}.
From the data of the quiver, we find that the corresponding 5d SCFT has the following gauge and flavor ranks:
\begin{equation}
    r = 2 \, , \quad  f = 1 \, .
\end{equation}

In this orbifold of the Laufer singularity, the $\Z_3$ quotient for $k=2$, we have that the theory has a non trivial $\Z_3$ electro-magnetic higher form symmetry, as can be readily computed from the BPS quiver in Figure \ref{fig:lauferZ3 quiver}. As it is easy to check, the $\Z_3$ orbifold acts without fixed loci, showing that the BPS computation agrees with the general algebraic expectation.

\subsection{Orbifolds of orbifolds}
In this Section, we exhibit an example of an abelian orbifold of a non-abelian orbifold of $\mathbb{C}^3$. Namely, our starting point is a threefold obtained as:
\begin{equation}
   X = \mathbb{C}^3/\Gamma, 
\end{equation}
with $\Gamma \in SL(3,\mathbb{C})$. If $\Gamma$ is non-abelian then $X$ is non-toric. This class of orbifolds was studied in \cite{DelZotto:2022fnw,Tian:2021cif}, in which the features of the corresponding 5d SCFTs were carefully explored. Here, we wish to pursue our program by acting with a further \textit{abelian} orbifold on $X$.

\subsubsection{Non-abelian orbifolds with higher form symmetries}

The higher form symmetries of non-abelian orbifolds of $\mathbb{C}^3$ were classified in \cite{DelZotto:2022fnw}. Here we show in one example how, once again, one can identify the higher form symmetry group as a subgroup of the quantum dual and unorbifold it to get a theory without center symmetries.

Let us consider the case of the $D_{n,q}$ theories. We can fix $n=4, q=2$ for sake of simplicity. This group is generated by
\begin{gather}
    D_{n,q}=\Bigg\langle
        \begin{pmatrix}
            i & 0 & 0\\
            0 & -i & 0\\
            0 & 0 & 1
        \end{pmatrix},
        \begin{pmatrix}
            0 & i & 0 \\
            i & 0 & 0 \\
            0 & 0 & 1
        \end{pmatrix}\cdot
        \begin{pmatrix}
            e^{\pi i/4} & 0 & 0\\
            0 & e^{\pi i/4} & 0\\
            0 & 0 & e^{-\pi i/2}
        \end{pmatrix}
    \Bigg\rangle.
\end{gather}

and the character table is
\begin{align}\begin{matrix}
    \rho_1 \\
    \rho_2 \\
    \rho_3 \\
    \rho_4 \\
    \rho_5 \\
    \rho_6 \\
    \rho_7 \\
    \rho_8 \\
    \rho_9 \\
    \rho_{10} \\
\end{matrix}
\begin{pmatrix}
1 & 1 & 1 & 1 & 1 & 1 & 1 & 1 & 1 & 1 \\
1 & 1 & 1 & 1 & -1 & -1 & 1 & -1 & -1 & 1 \\
1 & 1 & 1 & 1 & -1 & -1 & -1 & 1 & 1 & -1 \\
1 & 1 & 1 & 1 & 1 & 1 & -1 & -1 & -1 & -1 \\
1 & -1 & -1 & 1 & -1 & 1 & i & i & -i & -i \\
1 & -1 & -1 & 1 & -1 & 1 & -i & -i & i & i \\
1 & -1 & -1 & 1 & 1 & -1 & -i & i & -i & i \\
1 & -1 & -1 & 1 & 1 & -1 & i & -i & i & -i \\
2 & 2 i & -2 i & -2 & 0 & 0 & 0 & 0 & 0 & 0 \\
2 & -2 i & 2 i & -2 & 0 & 0 & 0 & 0 & 0 & 0
\end{pmatrix}
\end{align}
from which we have both the quiver and the quantum dual as well as the higher form symmetry generator. The quiver is depicted in Figure \ref{fig: Dn,q}, where we label the quiver nodes by the corresponding representations.

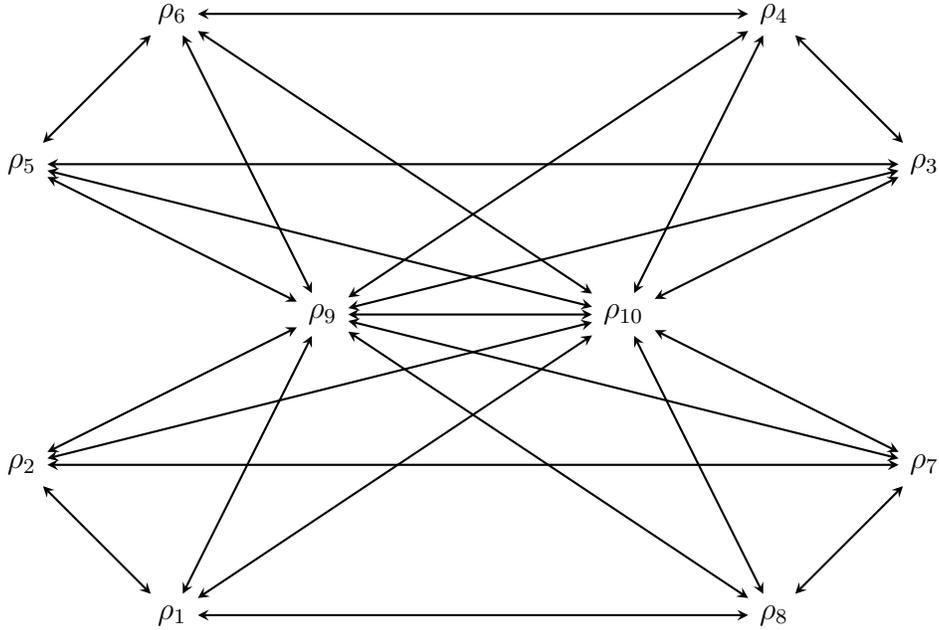
\begin{figure}[H]
\begin{center}
\scalemath{1}{
\begin{tikzpicture}[>=stealth, thick]
% Nodes
\node (M_1) at (-2,0) {$\rho_9$};
\node (M_2) at (2,0) {$\rho_{10}$};
\node (M_3) at (6,2) {$\rho_3$};
\node (M_4) at (4,4) {$\rho_4$};
\node (M_5) at (-6,2) {$\rho_5$};
\node (M_6) at (-4,4) {$\rho_6$};
\node (M_7) at (6,-2) {$\rho_7$};
\node (M_8) at (4,-4) {$\rho_8$};
\node (M_9) at (-6,-2) {$\rho_2$};
\node (M_{10}) at (-4,-4) {$\rho_1$};
%%%%
\draw[<->] (M_1) to node[above,yshift = -3pt]{} (M_2);
\draw[<->] (M_1) to node[above]{} (M_3);
\draw[<->] (M_1) to node[above]{} (M_4);
\draw[<->] (M_1) to node[above]{} (M_5);
\draw[<->] (M_1) to node[above]{} (M_6);
\draw[<->] (M_1) to node[above]{} (M_7);
\draw[<->] (M_1) to node[above]{} (M_8);
\draw[<->] (M_1) to node[above]{} (M_9);
\draw[<->] (M_1) to node[above]{} (M_{10});
\draw[<->] (M_2) to node[above]{} (M_3);
\draw[<->] (M_2) to node[above]{} (M_4);
\draw[<->] (M_2) to node[above]{} (M_5);
\draw[<->] (M_2) to node[above]{} (M_6);
\draw[<->] (M_2) to node[above]{} (M_7);
\draw[<->] (M_2) to node[above]{} (M_8);
\draw[<->] (M_2) to node[above]{} (M_9);
\draw[<->] (M_2) to node[above]{} (M_{10});
\draw[<->] (M_3) to node[right]{} (M_4);
\draw[<->] (M_5) to node[left]{} (M_6);
\draw[<->] (M_7) to node[right]{} (M_8);
\draw[<->] (M_9) to node[left]{} (M_{10});
\draw[<->] (M_6) to node[left]{} (M_4);
\draw[<->] (M_5) to node[left]{} (M_3);
\draw[<->] (M_9) to node[left]{} (M_7);
\draw[<->] (M_{10}) to node[left]{} (M_8);
\end{tikzpicture}
}
\end{center}
\caption{BPS quiver for the geometry $\mathbb{C}^3/D_{4,2}$.}
\label{fig: Dn,q}
\end{figure}

The quantum dual is the $\mathbb{Z}_4 \times \mathbb{Z}_2$ group generated by the eight 1d irreps of the group and the irreps generating the higher form symmetry subgroups are the trivial ones corresponding to $\rho_1$ and $\rho_4$ (the fourth row trivializes the group elements that act with fixed loci).

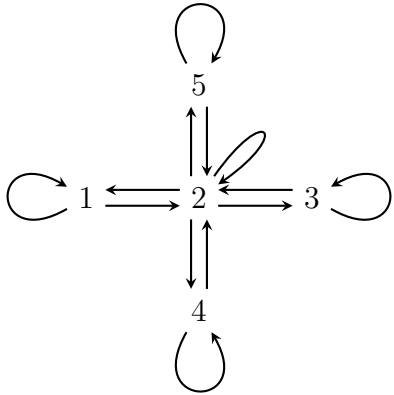
\begin{figure}[H]
\begin{center}
\begin{tikzpicture}[>=stealth, thick]
% Nodes
\node (1) at (-1.5,0) {$1$};
\node (2) at (0,0) {$2$};
\node (3) at (1.5,0) {$3$};
\node (4) at (0,-1.5) {$4$};
\node (5) at (0,1.5) {$5$};
% Loops
\draw[->] (1) edge[->, min distance=12mm, in=150, out=-150] node[left]{} (1);
\draw[->] (2) edge[->, min distance=12mm, in=35, out=55] node[right]{} (2);
\draw[->] (3) edge[->, min distance=12mm, in=30, out=-30] node[right]{} (3);
\draw[->] (4) edge[->, min distance=12mm, in=-60, out=-120] node[left]{} (4);
\draw[->] (5) edge[->, min distance=12mm, in=60, out=120] node[left]{} (5);
%%%%%
\draw[<-, transform canvas={yshift=3pt}] (1)--(2)   {};
\draw[<-, transform canvas={yshift=-3pt}] (2)--(1)  {};
\draw[<-, transform canvas={yshift=3pt}] (2)--(3)   {};
\draw[<-, transform canvas={yshift=-3pt}] (3)--(2)  {};
\draw[<-, transform canvas={xshift=3pt}] (2)--(5)   {};
\draw[<-, transform canvas={xshift=-3pt}] (5)--(2)  {};
\draw[<-, transform canvas={xshift=3pt}] (2)--(4)   {};
\draw[<-, transform canvas={xshift=-3pt}] (4)--(2)  {};
\end{tikzpicture}
\end{center}
    \caption{BPS quiver for the $D_4\times \mathbb{C}$ threefold.}
    \label{fig:D4affine}
\end{figure}

We can now proceed in taking the quotient with respect to the Ab$(\Gamma/N)$ by identifying the nodes in the orbit of $\rho_4$ giving us the quiver in Figure \ref{fig:D4affine}.
The quiver indeed shows that the theory is free of higher form symmetries.

One could expect such result by looking at the generators and noticing that the abelian factor acting on all coordinates commutes with the $SU(2)$ subgroup that acts on the first two. This is somehow analogous to the toric case described in \cref{sec:orbifolding}, where this time the starting point is the affine $D_4$ quiver.

\subsubsection{Non-abelian orbifolds without higher form symmetries}

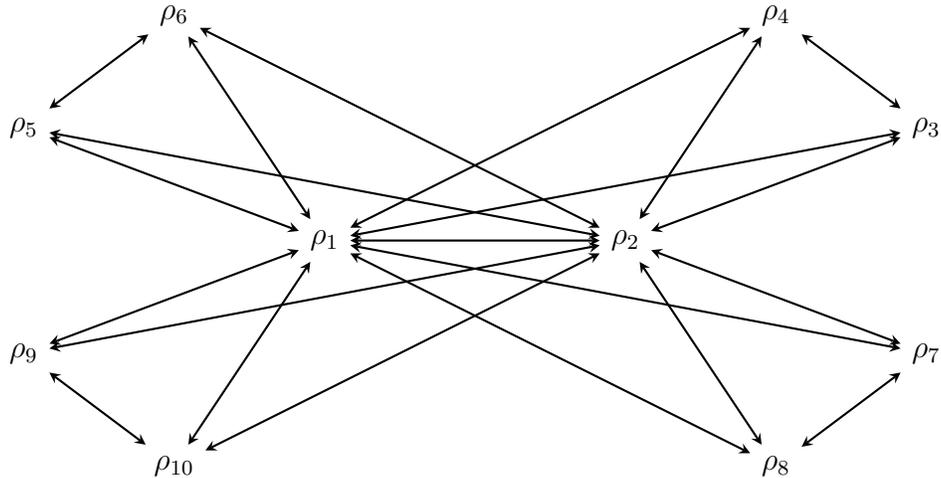
\begin{figure}[H]
\begin{center}
\scalemath{1}{
\begin{tikzpicture}[>=stealth, thick]
% Nodes
\node (M_1) at (-2,0) {$\rho_1$};
\node (M_2) at (2,0) {$\rho_2$};
\node (M_3) at (6,1.5) {$\rho_3$};
\node (M_4) at (4,3) {$\rho_4$};
\node (M_5) at (-6,1.5) {$\rho_5$};
\node (M_6) at (-4,3) {$\rho_6$};
\node (M_7) at (6,-1.5) {$\rho_7$};
\node (M_8) at (4,-3) {$\rho_8$};
\node (M_9) at (-6,-1.5) {$\rho_9$};
\node (M_{10}) at (-4,-3) {$\rho_{10}$};
%%%%
\draw[<->] (M_1) to node[above,yshift = -3pt]{} (M_2);
\draw[<->] (M_1) to node[above]{} (M_3);
\draw[<->] (M_1) to node[above]{} (M_4);
\draw[<->] (M_1) to node[above]{} (M_5);
\draw[<->] (M_1) to node[above]{} (M_6);
\draw[<->] (M_1) to node[above]{} (M_7);
\draw[<->] (M_1) to node[above]{} (M_8);
\draw[<->] (M_1) to node[above]{} (M_9);
\draw[<->] (M_1) to node[above]{} (M_{10});
\draw[<->] (M_2) to node[above]{} (M_3);
\draw[<->] (M_2) to node[above]{} (M_4);
\draw[<->] (M_2) to node[above]{} (M_5);
\draw[<->] (M_2) to node[above]{} (M_6);
\draw[<->] (M_2) to node[above]{} (M_7);
\draw[<->] (M_2) to node[above]{} (M_8);
\draw[<->] (M_2) to node[above]{} (M_9);
\draw[<->] (M_2) to node[above]{} (M_{10});
\draw[<->] (M_3) to node[right]{} (M_4);
\draw[<->] (M_5) to node[left]{} (M_6);
\draw[<->] (M_7) to node[right]{} (M_8);
\draw[<->] (M_9) to node[left]{} (M_{10});
\end{tikzpicture}
}
\end{center}
\caption{BPS quiver for the geometry in \cref{X orbifold} for $m=2$.}
\label{fig: G2 quiver}
\end{figure}

We can also consider the case of a theory without higher form symmetries and perform a gauging that introduces them. Let us first define $X$ as the quotient of $\mathbb{C}^3$ by the group $G_m$. The generators of $G_m$ are:
\begin{equation}
    G_m = \Bigg\langle\begin{pmatrix}
       \xi & 0 & 0 \\
       0 & \xi^{-1} & 0 \\
       0 & 0 & 1 \\ 
    \end{pmatrix},\begin{pmatrix}
       -1 & 0 & 0 \\
       0 & 1 & 0 \\
       0 & 0 & -1 \\ 
    \end{pmatrix},\begin{pmatrix}
       0 & 1 & 0 \\
       1 & 0 & 0 \\
       0 & 0 & -1 \\ 
    \end{pmatrix} \Bigg\rangle,
\end{equation}
with $\xi^{2m}=1$, acting on $\mathbb{C}[x,y,z]$ in the usual way. The invariant coordinates under the action are:
\begin{equation}
    X = z^2, \quad Y = x^2y^2, \quad Z = x^{2m}+y^{2m}, \quad  W = xyz(x^{2m}-y^{2m}).
\end{equation}
There is a unique relation between these invariants, and therefore the threefold $X = \mathbb{C}^3/G_{m}$ is a hypersurface in $\mathbb{C}^4$:
\begin{equation}\label{X orbifold}
 X:\quad   W^2 -XYZ^2 + 4XY^{m+1} = 0, \subset \mathbb{C}^4.
\end{equation}
The associated BPS quiver, e.g.\ for the case $m=2$, is shown in Figure \ref{fig: G2 quiver}.

Notice that for $m = 2k$ the orbifold \cref{X orbifold} enjoys the following action, that preserves the threefold and the holomorphic 3-form:
\begin{equation}\label{X isometry}
    (X,Y,Z,W) \rightarrow (\lambda X,\lambda Y,\lambda Z, \lambda W), \quad \lambda \in \mathbb{Z}_2,
\end{equation}
which can be achieved at the level of the original $\mathbb{C}^3$ coordinates by extending the group action by $\mathbb{Z}_2$ adding the following generator
\begin{equation}
    G_m^{ext} = \Bigg\langle G_m,\begin{pmatrix}
       e^{\pi i/4} & 0 & 0 \\
       0 & e^{\pi i/4} & 0 \\
       0 & 0 & e^{-\pi i/2} \\ 
    \end{pmatrix} \Bigg\rangle,
\end{equation}

\begin{figure}[H]
\begin{center}
\scalemath{0.5}{
\begin{tikzpicture}[>=stealth, thick]
% Nodes
\node (M_1) at (0,2) {$\rho_1$};
\node (M_2) at (0,-2) {$\rho_2$};
\node (M_3) at (4,0) {$\rho_3$};
\node (M_4) at (-4,0) {$\rho_4$};
\node (M_5) at (16,2) {$\rho_5$};
\node (M_6) at (-16,2) {$\rho_6$};
\node (M_7) at (8,3) {$\rho_7$};
\node (M_8) at (-8,3) {$\rho_8$};
\node (M_9) at (12,4.5) {$\rho_9$};
\node (M_{10}) at (-12,4.5) {$\rho_{10}$};
\node (M_{11}) at (4,6) {$\rho_{11}$};
\node (M_{12}) at (-4,6) {$\rho_{12}$};
\node (M_{13}) at (16,-2) {$\rho_{13}$};
\node (M_{14}) at (-16,-2) {$\rho_{14}$};
\node (M_{15}) at (8,-3) {$\rho_{15}$};
\node (M_{16}) at (-8,-3) {$\rho_{16}$};
\node (M_{17}) at (12,-4.5) {$\rho_{17}$};
\node (M_{18}) at (-12,-4.5) {$\rho_{18}$};
\node (M_{19}) at (4,-6) {$\rho_{19}$};
\node (M_{20}) at (-4,-6) {$\rho_{20}$};
%%%%
\draw[->] (M_{11}) to  (M_9);
\draw[->] (M_9) to  (M_5);
\draw[->] (M_5) to  (M_7);
\draw[->] (M_7) to  (M_{11});
%%%
\draw[->] (M_{10}) to  (M_{12});
\draw[->] (M_{12}) to  (M_8);
\draw[->] (M_8) to  (M_6);
\draw[->] (M_6) to  (M_{10});
%%%%
\draw[->] (M_{14}) to  (M_{16});
\draw[->] (M_{16}) to  (M_{20});
\draw[->] (M_{20}) to  (M_{18});
\draw[->] (M_{18}) to  (M_{14});
%%%
\draw[->] (M_{15}) to  (M_{13});
\draw[->] (M_{13}) to  (M_{17});
\draw[->] (M_{17}) to  (M_{19});
\draw[->] (M_{19}) to  (M_{15});
%%%%
\draw[->] (M_1) to  (M_3);
\draw[->] (M_3) to  (M_2);
\draw[->] (M_2) to  (M_4);
\draw[->] (M_4) to  (M_1);
%%%%
\draw[->] (M_{12}) to  (M_1);
\draw[->] (M_9) to  (M_1);
\draw[->] (M_{13}) to  (M_1);
\draw[->] (M_{16}) to  (M_1);
\draw[<-] (M_{10}) to  (M_1);
\draw[<-] (M_{11}) to  (M_1);
\draw[<-] (M_{15}) to  (M_1);
\draw[<-] (M_{14}) to  (M_1);
%%%%%
\draw[->] (M_{19}) to  (M_2);
\draw[->] (M_{18}) to  (M_2);
\draw[->] (M_6) to  (M_2);
\draw[->] (M_7) to  (M_2);
\draw[<-] (M_{20}) to  (M_2);
\draw[<-] (M_{17}) to  (M_2);
\draw[<-] (M_8) to  (M_2);
\draw[<-] (M_5) to  (M_2);
%%%%%
\draw[->] (M_{10}) to  (M_4);
\draw[->] (M_{14}) to  (M_4);
\draw[->] (M_{15}) to  (M_4);
\draw[->] (M_{11}) to  (M_4);
\draw[<-] (M_{18}) to  (M_4);
\draw[<-] (M_6) to  (M_4);
\draw[<-] (M_7) to  (M_4);
\draw[<-] (M_{19}) to  (M_4);
%%%%%
\draw[->] (M_5) to  (M_3);
\draw[->] (M_{17}) to  (M_3);
\draw[->] (M_{20}) to  (M_3);
\draw[->] (M_8) to  (M_3);
\draw[<-] (M_{13}) to  (M_3);
\draw[<-] (M_9) to  (M_3);
\draw[<-] (M_{16}) to  (M_3);
\draw[<-] (M_{12}) to  (M_3);
\end{tikzpicture}
}
\end{center}
\caption{BPS quiver for the geometry in \cref{G2 orbifold} for $k=1$.}
\label{fig: G2 quiver orbifold}
\end{figure}
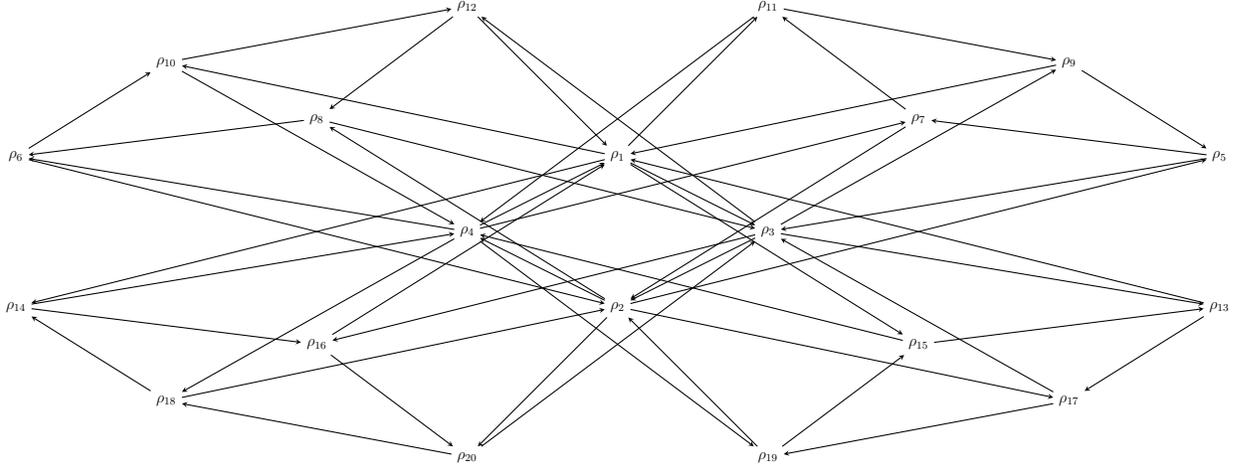

Note that this extra generator, despite being of order eight, $g^8 = 1$, acts effectively as an order 2 symmetry of the above polynomial, since $g^2$ can be expressed in terms of the generators of $G_{2k}$. We can now apply the usual McKay correspondence for the new group $G_{2k}^{exp}$ to get the BPS quiver (e.g.\ choosing $k=1$) shown in Figure \ref{fig: G2 quiver orbifold}.

One can check that now the torsional part of the cokernel is no longer trivial and the orbifolded theory has a $\mathbb{Z}_2$ electric/magnetic higher form symmetry. The BPS quiver for larger $k$ can be straighforwardly obtained employing our technology.

Orbifolding the initial threefold $X$ by the isometry in \cref{X isometry}, we obtain the following threefold, where the invariant coordinates $a_i$ can be obtained exactly as in \cref{eq:invariants F0}:
\begin{equation}\label{G2 orbifold}
\begin{cases}
\begin{array}{cccc}
a_1a_2 = a_5^2, \quad & a_3a_4 = a_{10}^2,\quad & a_5 a_8 = a_2 a_6,\quad & a_8 a_{10} = a_3 a_9,\\
a_1a_3 = a_6^2, \quad & a_5 a_{10} = a_6 a_9, \quad & a_5a_9 = a_2a_6, \quad & a_7 a_9 = a_4 a_5,\\
a_1a_4 = a_7^2, \quad & a_5a_{10} = a_7a_8, \quad & a_6a_7 = a_1a_{10}, \quad & a_7a_{10} = a_4a_6,\\
a_2a_3 = a_8^2, \quad & a_5a_6 = a_1a_8, \quad & a_6a_8 = a_3a_5, \quad & a_8 a_9 = a_2a_{10},\\
a_2a_4 = a_9^2 ,\quad & a_5a_7 = a_1a_0, \quad & a_6a_{10} = a_3a_7, \quad & a_9a_{10} = a_4a_8,\\
\boxed{a_4 -a_5a_3 +4a_5 a_2^k = 0 } .\quad &&&\\
\end{array}
\end{cases}
\end{equation}
The boxed equation is the orbifold-invariant version of \cref{X orbifold}, that plays the role of a non-toric linear constraint. 
The threefold \cref{G2 orbifold} is a non-toric non-complete intersection, that engineers a 5d SCFT with:
\begin{equation}
    r =6 \, , \quad f = 7 \, .
\end{equation}

\section{Conclusions and outlook}\label{sec:conclusions}

In this work we have explored the relation between orbifolds and higher form symmetries in the context of geometric engineering. The main result is that in 5d SCFTs engineered via M-theory on both toric and non-toric CY threefolds constructed as orbifolds of some parent CY$_3$, the higher form symmetries of the theory appear as a symmetry of the BPS quiver in a specific mutant. In particular, the Defect Group can be put into correspondence with a subgroup of the quantum symmetry of the orbifolded quiver. Using mathematical results described in the appendices, we were able to show that the (un-)orbifolding procedure can be carried out in full generality for any quiver and showed in many (non-)toric examples the orbifold origin of higher symmetries.

We have explicitly computed the singular threefold geometry after the orbifold operation, examining a hefty list of 5d SCFTs which are engineered by M-theory on non-toric non-complete intersection CY$_3$. It is clear that these orbifolding constructions can be straighforwardly generalized to any canonical Calabi-Yau threefold that possesses a known presentation in terms of algebraic equations in some affine space. Our procedure will then predict what is the defect group of the resulting theory. This endeavour, which we postpone to future work, will allow us to scan the landscape of non-toric non-complete intersection singularities, which are thought to comprise the large majority of canonical CY$_3$.\footnote{For some recent examples of 5d SCFTs engineered via such exotic canonical CY$_3$, see \cite{trinions}.}

Aside from the geometrical considerations, we mostly focused on the BPS quiver analysis, and in particular on cyclic and finite abelian orbifolds. A natural question is how this operation can be generalized to the case of non-abelian orbifolds. We are currently working on a full classification of all possible orbifolds of the conifold, both abelian and non-abelian, to provide insights on the possible relation between the geometrical orbifold operation and related physical quantities, such as the defect group. In particular, as mentioned in \cref{sec:quantumdual}, we wish to investigate the interplay between group extensions of the orbifold and higher group structures in the field theory.

Another direction which we have not explored in the current work is the use of Mayer-Vietoris' sequences, as in \cite{Cvetic:2022imb, Cvetic:2024dzu}, to directly study the boundary geometry of the Sasaki-Einstein base of the threefold $X$. The torsional part of $H_{1}(\partial X)$ encodes the higher form symmetry of the engineered theory, however $H_{3}(\partial X)$ seems to be sensible to the full quantum symmetry. Clarifying this relation might help establish a general criterion to fully un-orbifold a singularity, providing a precise notion of “minimality'' of a theory, a definition that might help in the classification program for 5d SCFTs.

Recently, the same group published a paper relating the anomalies for higher form symmetries to APS invariants \cite{Cvetic:2025lat}. Their analysis relies on a choice of a certain twist of the Dirac operator on the engineering geometry that can be associated to the higher form symmetry. This twist can be related to the quantum symmetry group of the quivers we describe in this work. Some generalization of their work and its relation to the BPS quiver is currently a work in progress \cite{eta_from_quiver}.

As a further future direction, we would like to broaden our reasoning to $G_2$ manifolds as well. These manifolds are obtained by considering suitable orbifold fibrations, providing examples for the Defect group/quantum group correspondence outside the context of BPS quivers.

Finally, we conclude noticing that our analysis was able to spot higher group structures in the field theory, using the relation between quantum symmetry and structure group. It is natural to ask to which extent this connection holds and how much of the flavour symmetry of a field theory can be led back to orbifold operations. Recently \cite{Dierigl:2023jdp, DeMarco:2025pza, Chakrabhavi:2025bfi, Acharya:2025dty}, space-time symmetries such as charge conjugation and $SL_2(\Z)$ duality symmetry in 5d SCFT where studied in the context of Geometric Engineering, and it is worth asking if the BPS quiver can detect them.

\section*{Acknowledgments}
The authors would like to thank Jonathan J. Heckman and Max H\"ubner for insightful comments on the draft. AS and DD wish to thank Marina Moleti for an interesting discussion. The authors would also like to thank Andrés Collinucci, Fabrizio Del Monte, Mario De Marco, Marina Moleti and Roberto Valandro for valuable comments on the draft and for coordinating the submission of their work. SNM would like to thank Vivek Chakrabhavi, Mirjam Cveti\v{c}, Jonathan J. Heckman and Max H\"ubner for discerning conversations about this work and on related topics. The research of AS is funded from the VR Centre for Geometry and Physics (VR grant No.\ 2022-06593) at Uppsala University. AS also acknowledges the ERC grant No. 101171852-HIGH and the VR project grant No. 2023-05590 for partial support. The work of SNM is supported by DOE (HEP) Award DE-SC0013528 as well as by BSF grant 2022100 and in part by a University Research Foundation grant at the University of Pennsylvania. The work of DD is funded by the Swedish Foundation for International Cooperation in Research and Higher Education (STINT).

\clearpage

\appendix

\section{Skew-group constructions}\label{app:skew_group}
In this appendix, we provide mathematical details explaining how to obtain new quivers from old quivers with a group action, and how to reverse this operation. This goes back to the work of Reiten and Riedtmann in \cite{ReitenRiedtmann:1985}, and we provide further references along the way. 

We begin with a reminder on skew-group algebras. For this, let $A$ be a (associative, unital) $\mathbb{C}$-algebra, and let the group $G$ act (from the left) on $A$ via $\mathbb{C}$-algebra automorphisms, i.e.\ we write the action of $g \in G$ on an element $a \in A$ as $g(a)$.

The \emph{skew-group algebra} $A \ast G$ of $A$ by the action of $G$ then is
\[ A \ast G = A \otimes \mathbb{C}G, \]
with multiplication defined by
\[ (a \otimes g) \cdot  (b \otimes h) = (a g(b) \otimes gh).  \]

\subsection{Skewed path algebras}
We summarise the results of Demonet in \cite{Demonet:2010}. Let $A = \mathbb{C}Q$ be the path algebra of a (finite) quiver $Q = (Q_0, Q_1)$, and let $G$ be a finite group acting on $A$. From now on, we assume that $G$ acts faithfully. Under some further mild assumptions on the action, one can compute the quiver of the skew-group algebra $A \ast G$, which we denote $Q \ast G$ and call the \emph{skewed quiver}. 

More precisely, we assume that $G$ acts on $A$ such that $G$ permutes the idempotents corresponding to $Q_0$, and such that the arrow space $\mathbb{C} Q_1$ is stabilised by $G$. Note that this means that $G$ can map an arrow to a linear combination of arrows. 

We set up some notation first. By assumption, we can identify the action of $G$ on the vertices in $Q_0$ with the action on the corresponding idempotents, so we write $Q_0 = \{ 1, \ldots, m\}$ and work with orbits on $Q_0$. The set of orbits is denoted $Q_0/G$, and for a vertex $i$, its orbit is written $[i]$, and its stabiliser is $G_i$. For two vertices, we denote the intersection of stabilisers by $G_{i,j} = G_i \cap G_j$. Next, we fix a transversal $T$ for the action of $G$ on $Q_0$, and denote the element in $T$ corrsponding to $[i]$ by $t_i$. Furthermore, for each $i$, we fix an element $g_i \in G$ such that $g_i(t_i) = i$. 
Next, note that $G$ also acts on the product of orbits $[i] \times [j]$. We denote by $O_{i,j}$ a transversal for this action. 
Finally, we denote by $A_{i,j}$ the vector space spanned by arrows from $i$ to $j$. 

Demonet then constructs the quiver $Q \ast G$ as follows. The vertices are 
\[ (Q \ast G)_0 = \bigcup_{[i] \in Q_0/G} \{ [i] \} \times \operatorname{Irr}(G_i), \]
where $\operatorname{Irr}(G_i)$ denotes the set of irreducible representations of $G_i$. The set of arrows from $([i], \chi)$ to $([j], \tau)$ is given by any basis of 
\[ \bigoplus_{(i',j') \in O_{i,j} } \operatorname{Hom}_{G_{i',j'}} ( (\chi^{g_{i'}})_{|G_{i', j'} } , (\tau^{g_{j'}})_{|G_{i', j'}} \otimes_{\mathbb{C}} A_{i',j'}  ).     \]
Here, the representation $\chi^{g_{i'}}$ is the representation of $G_{i'}$ given by $h \mapsto \chi( h^{g_{i'}^{-1}} )$.
Demonet then shows in \cite[Theorem 1]{Demonet:2010} that under the above assumptions, we have a Morita equivalence
\[ (\mathbb{C} Q) \ast G \simeq_M \mathbb{C} (Q \ast G). \]

Next, suppose that $A = \mathbb{C} Q/I$ is a quotient of a path-algebra by an ideal $I$ contained in the square of the arrow ideal of $\mathbb{C}Q$. Suppose further that the $G$ on $\mathbb{C} Q$ stabilises $I$, then the action descends to $\mathbb{C} Q/I$, and $A \ast G$ is Morita equivalent to a quotient of $\mathbb{C} (Q \ast G)$. 

We now make some remarks on how this procedure reproduces some well-known results as special cases. Firstly, if the original quiver $Q$ has a single vertex, then we may view the arrow space as a faithful representation $\rho = \mathbb{C} Q_1 $  of the finite group $G$. To construct $Q \ast G$, we consider the single orbit, which is the single vertex, and make copies of it indexed by the irreducible representations of $G$, since $G$ is the stabiliser of the single vertex. We then draw arrows between vertices corresponding to irreducible representations according to bases of the Hom-spaces $\operatorname{Hom}_G ( \chi, \tau \otimes \rho)$. This simply recovers the McKay quiver of $G$ with respect to the representation $\rho$. 
At the other extreme, if instead $G$ acts freely on the set of vertices and the \emph{set} of arrows, all involved stabilisers of vertices are trivial. To obtain the vertices of $Q \ast G$, we collapse all vertices to a single orbit. Since the stabilisers are trivial, the relevant Hom-spaces are representations of the trivial group, i.e.\ simply vector spaces. By counting dimensions, we see that the quiver $Q \ast G$ can be obtained as the set-theoretic quotient $(Q_0/G, Q_1/G)$. 
Further simplifications of the above procedure are possible depending on the properties of the action. 

\subsection{Unskewing}
It is a useful feature of the skewing procedure that it can be reversed by another skewing, so long as the group $G$ is abelian. This is proved by Reiten and Riedtmann in \cite{ReitenRiedtmann:1985}. However, the same result follows from Cohen-Montgomery duality, which may be of independent interest, so we follow this approach. To phrase this duality, we remind the reader that the skew-group algebra $A \ast G$ is a special case of a so-called Hopf smash-product. For this, let $H$ be a Hopf-algebra and $A$ an $H$-module algebra. Relevant for us is the case when $H = \mathbb{C} G$ is a group algebra, and in this case all possible such $H$-module structures arise from $G$-actions, and vice versa. Furthermore, the smash-product is simply the skew-group algebra $A \sharp H \simeq A \ast G$. 
The so-called Cohen-Montgomery duality \cite[Theorem 3.2]{Cohen:1984group} then states that, up to Morita equivalence, we can recover $A$ from $A \sharp H$ by smashing with the dual Hopf algebra $H^\ast$, i.e.\ we have 
\[ A \simeq_M (A \sharp H ) \sharp H^\ast. \]

The reader may notice that for $H = \mathbb{C} G$, the dual $H^\ast$ is in general not a group algebra. This is precisely the case only when $G$ is abelian, as in this case $H^\ast = \mathbb{C} \hat{G}$ where $\hat{G } = \operatorname{Hom}(G, \mathbb{C}^\ast)$ is the dual group. Thus, in this situation we can skew by another group action to recover $A$, that is when $G$ is abelian, we have 
\[ A \simeq_M (A \ast G) \ast \hat{G}. \]

Furthermore, it is clear that one can iterate this procedure along a composition series of $G$ with abelian factor groups. That means, for a solvable group $G$, we can recover $A$ from $A \ast G$ by performing a finite sequence of skews by abelian groups. See \cite{ReitenRiedtmann:1985} for details. 

\subsection{Iterated skewing and group extensions}\label{app:skew_group2}
Here, we discuss how group extensions can be used break down the skew of $A = \mathbb{C}Q$ by $G$ into several smaller skew-operations. 

We begin with the case when the group is a semi-direct product $G = N \rtimes H$. The first step is easy, if $G$ acts faithfully on $A = \mathbb{C} Q$, then so does the subgroup $N$, and we can form 
\[ A' = A \ast N. \]
It is intuitively clear that we would like to take 
\[ A \ast G \simeq (A \ast N) \ast G/N, \]
but for this we need to specify the correct action of $G/N$ on $A \ast N$. However, since $G$ is a semi-direct product, we can realise the group $G/N \simeq H$ as a subgroup $H \leq G$. Then $H$ acts on $A$ by restriction, and on $\mathbb{C} N$ by conjugation. This combines to an action of $G/N$ on $A' = \mathbb{C}Q \ast N$, and the resulting skew-group algebra is indeed 
\[  A \ast G \simeq (A \ast N) \ast G/N. \]
In particular, this means we can compute the quiver as an iterated skewed quiver as 
\[ Q \ast G = (Q \ast N) \ast G/N. \]

When the group extension $1 \to N \to G \to H \to 1$ does not split, the situation is slightly more involved. We can still choose a set-theoretic section $s \colon H \to G$ with $s(1) = 1$, but this now gives rise to a $2$-cocycle $\alpha \colon H \times H \to N$. Taking the same approach as before, we then obtain a \emph{crossed} action of $H$ on $A \ast N$, where the crossed product is precisely specified by the $2$-cocycle $\alpha$. One can then take the more general crossed product 
\[ (A \ast N) \ast_\alpha H, \]
also called the twisted smash product, and obtain as before 
\[ A \ast G \simeq (A \ast N) \ast_\alpha G/N. \]

\subsection{Examples}
As an example, we study the conifold, over $\mathbb{C}$, given by the ring 
\[ R = \mathbb{C}[x,y,z,w]/(xy-zw). \]
We first need an explicit description of the quiver we associate to $R$. This is done by taking a noncommutative crepant resolution of $R$. For this, fix the ideal $I = (x,z) \trianglelefteq R$ and consider the algebra 
\[A = \operatorname{End}_R(R \oplus I). \]
To write down a quiver presentation of $A$, we first note that 
\[ \Hom_R(R, R) \simeq \mathbb{C}, \; \Hom_R(I,I) \simeq \mathbb{C}, \]
so we can treat the summands $R$ and $I$ as vertices of the quiver. 

Next, we need the arrows, which are given by $\Hom_R(R, I)$ and $\Hom_R(I,R)$. These spaces are easy to describe by hand. Since $R$ is free, we have essentially two homomorphism from $R$ to $I$, given by sending $1 \in R$ to $x$ or to $z$. The other direction is slightly less obvious, but one reasonable choice is to describe $\Hom_R(I,R)$ as generated by the inclusion morphism taking $x \mapsto x$ and $z \mapsto z$, and the ``permutation'' of it given by taking $x \mapsto w$ and $z \mapsto y$. The first one will be called $i$, and the second one $j$. This way, we obtain the full quiver as follows.
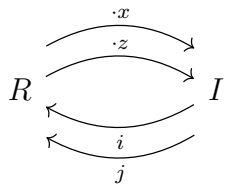
\[\begin{tikzcd}
R \arrow[rr, "\cdot x", bend left, shift left=4] \arrow[rr, "\cdot z", bend left] &  & I \arrow[ll, "j", bend left, shift left=4] \arrow[ll, "i", bend left]
\end{tikzcd}\]

\subsubsection{Interpreting group actions}
Next, we want to have a finite group acting on $R$, and extend this to an action on $A$. We will consider the same group acting in two different ways. 

\paragraph{The partial scaling action} 
For this action, we consider $G \simeq C_2$ of order $2$, acting on $R$ by scaling the variables $x$ and $y$ by $-1$. Clearly, this preserves the defining equation of $R$, so we have an actual action on $R$ and not just on the polynomial ring. 

To see this action on $A$, consider how this interacts with vertices and arrows. The vertices are stabilised, since $R$ is stable under this action, and the ideal $I$ is also stable under the action. Let us go through the arrows in detail: 
\begin{enumerate}
    \item The arrow $\cdot x$ gets scaled by $-1$. To see this, consider the fact that the nontrivial element $g \in G$ acts on the morphism 
    \[ 1 \mapsto x \] as 
    \[ (g(1) \mapsto g(x)) = (1 \mapsto -x),  \]
    which is the negative of the original arrow. 
    \item The arrow $\cdot z$ gets scaled by $1$. You can see it in the same way, noting that 
    \[ (g(1) \mapsto g(z)) = (1 \mapsto z). \]
    \item The arrow $i$ gets scaled by $1$. To see this, note that we essentially scale both the source and the target, so two minus signs cancel. The inclusion has the effect 
    \[ x \mapsto x  \text{ and } z \mapsto z, \]
    which is the same morphism as 
    \[ -x \mapsto -x  \text{ and } z \mapsto z. \]
    \item The arrow $j$ gets scaled by $-1$. The same observation about parities of signs shows that 
    \[ x \mapsto w  \text{ and } z \mapsto y, \]
    gets sent under $g$ to 
    \[ -x \mapsto w  \text{ and } z \mapsto -y, \]
    which is the same as $-j$. 
\end{enumerate}

\paragraph{The full diagonal action}
We contrast the previous action with the one where $G$ acts on $R$ by scaling \emph{all} variables by $-1$, in a ``full diagonal action''. Going through the same motions, we can see that this has the effect of scaling both arrows $R \to I$ by $-1$, while scaling both arrows $I \to R$ by $1$. 

\subsubsection{Skewing the quiver}
Now, we use the two actions to compute the skew group algebra $A \ast G$ with respect to those actions. We remind the reader that the standard notation only mentions $A$ and $G$, but we implicitly have a group action attached to this notation as well. 

\paragraph{The partial scaling action}
We begin with the partial scaling action and compute the $Q \ast G$ as follows. The vertices $R$ and $I$ are both stabilised by the $G$-action, so we make copies of $R$ and of $I$ indexed by $\operatorname{Irr}(G)$, the irreducible representations of the stabiliser, which happens to be all of $G$. Since there's only two irreducible representations of $G$, we label them by $1$ and $-1$. 

\[ \begin{tikzcd}
{(R, 1)}  &  & {(I,1)}   \\
          &  &           \\
{(R, -1)} &  & {(I, -1)}
\end{tikzcd}\]

To add in the arrows, we use the following computation. The arrows from $(R, \chi)$ to $(I, \tau)$ are given by 
\[ \Hom_{\mathbb{C}G}(\chi, \tau \otimes \rho_{R, I}), \]
where $\rho_{R, I}$ is the representation of $G$ afforded by the arrow space of arrows from $R$ to $I$. On the arrow space $R \to I$, we act by scaling one arrow by $-1$ and the other by $1$, hence we have a $2$-dimensional representation of $G$ which is the sum of the trivial and the non-trivial irreducible representation, hence we write $\rho_{R, I} = 1 \oplus (-1)$. 
Thus, the arrows $(R,1) \to (I, 1)$ are given by 
\begin{align*}
    \Hom_{\mathbb{C}G}(1, 1\otimes (1 \oplus (-1))) &= \Hom_{\mathbb{C}G}(1, 1 \oplus (-1)) \\
    &= \Hom_{\mathbb{C}G}(1, 1) \oplus \Hom_{\mathbb{C}G}(1, (-1)) = \mathbb{C} \oplus 0.
\end{align*} 
This means, we get one arrow $(R,1) \to (I, 1)$. 
Next, we take $(R,1) \to (I, -1)$, and get essentially the same computation:
\begin{align*}
    \Hom_{\mathbb{C}G}(1, (-1)\otimes (1 \oplus (-1))) &= \Hom_{\mathbb{C}G}(1, (-1) \oplus 1) \\
    &= \Hom_{\mathbb{C}G}(1, -1) \oplus \Hom_{\mathbb{C}G}(1, 1) = 0 \oplus \mathbb{C}
\end{align*}  
Again, we get one arrow. We fill in the picture to see what is happening: 
\[\begin{tikzcd}
{(R, 1)} \arrow[rr] \arrow[rrdd] &  & {(I,1)}   \\
                                 &  &           \\
{(R, -1)}                        &  & {(I, -1)}
\end{tikzcd}\]
We continue, this time taking $(R, -1) \to (I,1)$ and $(R,-1) \to (I,-1)$. In the same way as before, we get exactly one arrow, and hence our quiver looks like this.
\[\begin{tikzcd}
{(R, 1)} \arrow[rr] \arrow[rrdd]  &  & {(I,1)}   \\
                                  &  &           \\
{(R, -1)} \arrow[rruu] \arrow[rr] &  & {(I, -1)}
\end{tikzcd}\]

Now, we also need to look at arrows coming from $I$ and its copies. Luckily, the representation $\rho_{I,R}$ is the same as $\rho_{R, I}$. Note that we again scale one arrow by $1$ and one by $-1$. Thus, by symmetry we get the following quiver as the final result. 
\[\begin{tikzcd}
{(R, 1)} \arrow[rr, shift left=2] \arrow[rrdd, shift left=2]  &  & {(I,1)} \arrow[ll] \arrow[lldd]   \\
                                                              &  &                                   \\
{(R, -1)} \arrow[rruu, shift left=2] \arrow[rr, shift left=2] &  & {(I, -1)} \arrow[lluu] \arrow[ll]
\end{tikzcd}\]
This can be rearranged to a more familiar picture.  
\[\begin{tikzcd}
{(R, 1)} \arrow[rr, shift left=2] \arrow[dd, shift left=2] &  & {(I,1)} \arrow[ll] \arrow[dd]                               \\
                                                           &  &                                                             \\
{(I, -1)} \arrow[uu] \arrow[rr]                            &  & {(R, -1)} \arrow[uu, shift left=2] \arrow[ll, shift left=2]
\end{tikzcd}\]

\paragraph{The full diagonal action}
Now, we do the same procedure for the action that is scaling all the variables. The vertices are fixed as before, so we make the copies indexed by $\operatorname{Irr}(G)$. 
\[ \begin{tikzcd}
{(R, 1)}  &  & {(I,1)}   \\
          &  &           \\
{(R, -1)} &  & {(I, -1)}
\end{tikzcd}\]

The formulas are the same as before, but crucially now $\rho_{R,I}$ and $\rho_{I,R}$ are different. Recall that the arrows $R \to I$ both get scaled by $-1$, while the arrows $I \to R$ both get stabilised. This means, when looking at arrows $(R, 1) \to (I, 1)$ we have to compute 
\begin{align*}
    \Hom_{\mathbb{C}G}(1, 1\otimes ((-1) \oplus (-1))) &= \Hom_{\mathbb{C}G}(1, (-1) \oplus (-1))  \\
    &= \Hom_{\mathbb{C}G}(1, (-1)) \oplus \Hom_{\mathbb{C}G}(1, (-1)) = 0 \oplus 0.
\end{align*} 
This means there are no arrows from $(R, 1) \to (I, 1)$. 

Taking $(R, 1) \to (I, -1)$, we then get 
\begin{align*}
    \Hom_{\mathbb{C}G}(1, (-1)\otimes ((-1) \oplus (-1))) &= \Hom_{\mathbb{C}G}(1, 1 \oplus 1) \\
    &= \Hom_{\mathbb{C}G}(1, 1) \oplus \Hom_{\mathbb{C}G}(1, 1) = \mathbb{C} \oplus \mathbb{C},
\end{align*}   
so we get two arrows from $(R, 1) \to (I, -1)$. 

Continuing this computation, we get the following quiver
\[\begin{tikzcd}
{(R,1)} \arrow[rrdd, shift left] \arrow[rrdd, shift right]  &  & {(I,1)} \arrow[ll, shift left] \arrow[ll, shift right]  \\
                                                            &  &                                                         \\
{(R,-1)} \arrow[rruu, shift left] \arrow[rruu, shift right] &  & {(I,-1)} \arrow[ll, shift left] \arrow[ll, shift right]
\end{tikzcd}\]

\subsection{Application to orbifold singularities}
We now apply the above to the computation of quivers for orbifolds of possibly singular Calabi-Yau $3$-folds. We take the perspective of noncommutative crepant resolutions, and in particular caution the reader about a possible mismatch between actions on the $3$-fold and actions on the associated quiver. 

In the following, we assume that $X = \operatorname{Spec}(R)$ is an affine $3$-Calabi-Yau variety, and that $A = \operatorname{End}_R(M) \simeq \mathbb{C}Q/I$ is a noncommutative crepant resolution, presented by some quiver $Q$ and relations $I$. If the finite group $G$ acts on $X$ such that the orbifold $X/G$ stays Calabi-Yau, we would like to say that $Q \ast G$ is the quiver for $X/G$, and similarly if we instead start with an action of $G$ on $\mathbb{C}Q/I$ with the necessary homological assumptions, we would like to use $Q \ast G$ as the quiver for $X/G$. The situation is complicated by the fact that having a faithful action of $G$ on $X$ does not necessarily produce such an action on $A$, and vice versa. In the following situations, however, we can make the connection precise. 

If we start out with an action of $G$ on $X = \operatorname{Spec}(R)$, this is equivalent to an action on $R$, and we can consider whether $M$ becomes a $G$-equivariant $R$-module. If that is the case, we obtain an action of $G$ on $A = \operatorname{End}_R(M)$, and with the correct identification of irreducible summands of $M$ with vertices in $Q$, the action satisfies the necessary requirements to compute $Q \ast G$ as the quiver for $X/G$. 
Conversely, if we start out with an action of $G$ on $A$, this restricts to an action of $G$ on the center $Z(A) \supseteq R$. If the restriction to $R$ is faithful, we obtain again that $Q \ast G$ is the quiver for $X/G$.

\section{Toric diagrams and abelian orbifolds}\label{app:toric_theory}
In this section, we consider the case where $X$ is a Gorenstein toric variety, and $G$ is a finite abelian subgroup of the torus of $X$ with trivial determinant. We summarise how to describe the toric diagram of the orbifold $X/G$ given the diagram of $X$, and vice versa. All of this is standard in the toric literature, but usually phrased in more generality. We focus on the case when $X$ has dimension $3$, but of course the same holds in arbitrary dimension when interpreting the concept of a toric diagram correctly. We begin by recalling how a toric diagram fits into the standard language of polyhedral cones.

\begin{remark}
    Let $X$ be an affine Gorenstein toric variety of dimension $3$, with torus $T$, and $N = \Hom(\mathbb{C}^\ast, T) \simeq \mathbb{Z}^3$ the cocharacter lattice of $T$. The toric variety $X$ is defined by a lattice cone $\sigma \subseteq N \otimes_\mathbb{Z} \mathbb{R} = N_\mathbb{R}$. Crucially, since $X$ is Gorenstein, the cone $\sigma$ is in fact generated as the cone over a convex lattice polygon $P$. More precisely, after choosing coordinates for $N$, there exists a lattice polytope $P \subseteq \mathbb{Z}^2 \otimes_\mathbb{Z} \mathbb{R}$ such that $\sigma = \operatorname{Cone}(P \times \{ 1\} )$. This means, we can obtain $P$ from $\sigma$ by intersecting the cone with the hyperplane containing the primitive ray generators of $\sigma$ with respect to $N$, and return to the cone by adding an extra coordinate and taking the cone of this set. 
\end{remark}

Now, let $G \subseteq T$ be a finite subgroup of the torus of $X$. We want to take the quotient $X/G$, and describe this in terms of the lattice cone $\sigma$. Loosely speaking, the subgroup $G$ determines a second lattice $N' \subseteq N_\mathbb{R}$ containing $N$, such that $N'/N = \hat{G}$. The tori associated to $N$ and $N'$ then give an exact sequence 
\[  1 \to G \to T_N \to T_N', \]
and this extends to $X_{\sigma, N} = X$ such that 
\[ X_{\sigma, N'} =  X_{\sigma, N}/G. \]

The above is contained in \cite[Proposition 1.3.18]{Cox:2024toric}. In particular, $N'$ can be obtained explicitly by working in the dual lattice $N^\vee = M = \operatorname{Hom}(T, \mathbb{C}^\ast)$. Take $M_G = \{ \chi \in M \mid \chi(g) = 1 \text{ for all } g \in G\} $, the sublattice of characters vanishing on $G$, and then $(M_G)^\vee \supseteq M^\vee \simeq N$ gives the lattice $N'$. 

To see that the same procedure can be carried out one dimension lower with the toric diagrams, note the following. By assumption, we act with $G$ such that the quotient $X/G$ stays Gorenstein. To get $X/G$, we use the same cone $\sigma$, but with respect to a different lattice. 
Recall that the toric diagram for $X$ is the polygon $P$ given by the primitive ray generators of $\sigma$ with respect to $N$. In the same way, the toric diagram for $X/G$ is given by the polygon $P'$ given by the primitive ray generators of $\sigma$ with respect to $N'$. But since they both give rise to the same cone, they are similar polygons, and hence we obtain $P'$ from $P$ simply by changing the corresponding lattice in $\mathbb{R}^2$. This change of lattice can be computed directly from the similarity of the polygons. Furthermore, this can be simplified when $G$ acts crepantly. The action being crepant is equivalent to saying that the primitive ray generators of $\sigma$ stay the same with respect to $N$ or $N'$, so in this case it suffices to refine the lattice to go from $P$ to $P'$. 

We provide an example for this procedure. First, we consider as $X$ the conifold, with the action of the abelian group $\mathbb{Z}_{n_1} \times \mathbb{Z}_{n_2}$ as in \cref{eq:orbifold conifold}. In detail, we consider the $3$-torus $T$, with cocharacter lattice $N = \operatorname{Hom}(\mathbb{C}^\ast, T)$, with standard basis $(e_1, e_2, e_3)$. To build $X$, we choose the cone $\sigma$ over the toric diagram that is the square given by $(e_1, e_2, e_1 + e_3, e_2 + e_3)$. 
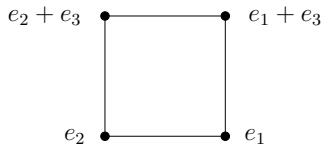
\begin{figure}[H]
\centering
    \scalebox{0.8}{
    \begin{tikzpicture}
        \draw (0,0)--(2,0)--(2,2)--(0,2)--(0,0); 
        
        \filldraw (0,0) circle (2pt);
        \filldraw (2,0) circle (2pt);
        \filldraw (2,2) circle (2pt);
        \filldraw (0,2) circle (2pt);

        \node at (-1,2) {$e_2 + e_3$};
        \node at (-0.5,0) {$e_2$};
        \node at (3,2) {$e_1 + e_3$};
        \node at (2.5,0) {$e_1$};
        
    \end{tikzpicture}}
    \caption{Toric diagram for the conifold}
    \end{figure}

More precisely, the above vectors are the primitive ray generators of $\sigma$. To avoid confusion, we denote the dual basis in $M = N^\vee$ by $(f_1, f_2, f_3)$. The dual cone is then generated by the vectors $(f_1, f_2, f_3, f_1 + f_2 -f_3)$, which gives us the presentation of the conifold as $\mathbb{C}[x_1, x_2, x_3, x_1 x_2 x_3^{-1}] \simeq \mathbb{C}[x,y,z,w]/(xy-zw)$. Furthermore, the torus $T$ acts as $(t_1, t_2, t_3) \mapsto (t_1, t_2, t_3, t_1 t_2 t_3^{-1})$. The group $G = \mathbb{Z}_{n_1} \times \mathbb{Z}_{n_2}$ by which we want to quotient sits inside the image of $T$ in $X$ as $(\lambda_1, \lambda_1^{-1}, \lambda_2, \lambda_2^{-1})$, and pulling this back to the coordinates of $T$ we have the subgroup $\langle (\lambda_1, \lambda_1^{-1}, 1), (1,1,\lambda_2) \rangle \leq T$. We now determine the lattice $N'$ as the dual of $M_G = \{ \chi \in M \mid \chi(g) = 1 \text{ for all } g \in G\}$. To describe $M_G$, we need to find all characters that vanish on $G$. This is simple because it suffices to find all characters that vanish on the two generators. Since $\lambda_1$ has order $n_1$ and $\lambda_2$ has order $n_2$, we have that $n_1 f_1, n_1 f_2, n_2 f_3$ are in $M_G$, and these are minimal with respect to each other. However, the fact that $\lambda_1$ acts ``diagonally'' means we have an extra vanishing character given by $f_1 + f_2$. To determine a basis for the lattice $M_G$, we can write down the matrix for these generators with respect to the basis $(f_i)$ as 
\[ \begin{pmatrix}
    n_1 & 0 & 1 & 0 \\ 0 & n_1 & 1 & 0 \\ 0 & 0 & 0 & n_2 
\end{pmatrix}. \]
Performing one step of integral Gaussian elimination produces the matrix 
\[ 
\begin{pmatrix}
n_1 & 0 & 1 & 0 \\ 0 & 0 & 1 & 0 \\ 0 & 0 & 0 & n_2    
\end{pmatrix},
\]
which reduces to the matrix 
\[ 
B = \begin{pmatrix}
n_1 &  1 & 0 \\ 0 &  1 & 0 \\ 0 &  0 & n_2    
\end{pmatrix}.
\]
Thus, the columns of $B$ form a basis for $M_G$. We then obtain a basis of the dual $N' = M_G^\vee$ by taking the rows of $B^{-1}$, which is given by 
\[
\begin{pmatrix}
    n_1^{-1} & - n_1^{-1} & 0 \\ 0 & 1 & 0 \\ 0 & 0 & n_2^{-1}
\end{pmatrix}.
\]
Finally, since $G$ acts crepantly on $X$, to obtain the toric diagram of $X/G$ from that of $X$, we simply keep the same square given by $(e_1, e_2, e_1 + e_3, e_2 + e_3)$, but now view it with respect to $N'$. We see that the vertical sides going from $e_1$ to $e_1 + e_3$ and from $e_2$ to $e_2 + e_3$ now have $n_2$ lattice points, since $N'$ contains the vector $\frac{1}{n_2} e_3$. Similarly, the vector $\frac{1}{n_1} e_1 - \frac{1}{n_1} e_2$ contributes $n_2$ many lattice points on the horizontal sides. Of course, combinations of these vectors produce a total of $n_1 \cdot n_2$ many lattice points in the square. This way, we obtain the diagram in Figure \ref{fig:conifold quiver}.

\section{Orbifolded theory via Chan-Paton fields}\label{app:chan_paton}

In this appendix we review the Chan-Paton construction for orbifolds. We will consider the example of $\mathbb{C}^3/\Gamma$, for $\Gamma \in SU(3)$ abelian, but the same argument generalizes to non-abelian orbifolds as well as other non-toric base spaces.

A D3-brane probing $\mathbb{C}^3/\Gamma$ has a world volume theory that can be obtained from the pure $\mathcal{N}=4$ SYM theory living on a single D3 probe on flat space by applying the following projections:
\begin{align}
    \gamma^{-1} A \gamma = A,  \quad \gamma^{-1} X \gamma = \omega^a X, \nonumber \\
    \gamma^{-1} Y \gamma = \omega^b Y,  \quad \gamma^{-1} Z \gamma = \omega^c Z,
\end{align}
where $A,X,Y,Z$ are generic $n\times n$ matrices corresponding to the fields of the D3 probe brane, $\gamma$ is a generator of $\Gamma$ in the regular representation. $\Gamma$ is taken to be of order $n$ and $\omega$ is the $n$-th root of unity acting on the coordinates of $\mathbb{C}^3$, with $\omega^{a+b+c}=1$, so that the resulting quotient geometry is Calabi-Yau.

As an explicit example let us consider the case $\mathbb{C}^3/\mathbb{Z}_3$ with action $(a,b,c)=(1,2,0)$. The system above becomes
\begin{align}
    \gamma^{-1} A \gamma = A,  \quad \gamma^{-1} X \gamma = e^{2 \pi i/3} X, \nonumber \\
    \gamma^{-1} Y \gamma = e^{4 \pi i/3} Y,  \quad \gamma^{-1} Z \gamma = Z,
\end{align}
with $\gamma = \text{diag}(1,e^{2 \pi i/3}, e^{4 \pi i/3} )$. The solutions to the above system are given by
\begin{align}\label{eq:orbi_mat}
    A = \begin{pmatrix} A_{11} & 0 & 0 \\ 0 & A_{22} & 0 \\ 0 & 0 & A_{33}\end{pmatrix},  \quad X = \begin{pmatrix} 0 & 0 & X_{13} \\ X_{21} & 0 & 0 \\ 0 & X_{32} & 0\end{pmatrix}, \nonumber \\
     Y = \begin{pmatrix} 0 & Y_{12} & 0 \\ 0 & 0 & Y_{23} \\ Y_{31} & 0 & 0\end{pmatrix},  \quad Z = \begin{pmatrix} Z_{11} & 0 & 0 \\ 0 & Z_{22} & 0 \\ 0 & 0 & Z_{33}\end{pmatrix}.
\end{align}
Each matrix entry represents a field in the bifundamental representation between two gauge groups $i \to j$. The superpotential is obtained by plugging these matrices in the $\mathcal{N}=4$ superpotential
\begin{align}
    W = \text{tr}(XYZ-YXZ) \, .
\end{align}

This procedure can now be iterated a second time, by considering the $A_{ij}/X_{ij}/Y_{ij}/Z_{ij}$ as matrices and requiring that they are projected in accordance to the geometric action. Let us consider the action:
\begin{align}
    \gamma^{-1} X_{32} \gamma = e^{2 \pi i/3} X_{32} \, , \quad  \gamma^{-1} X_{13} \gamma = e^{2 \pi i/3} X_{13} \, ,  \nonumber \\
    \gamma^{-1} Y_{12} \gamma = e^{2 \pi i/3} Y_{12} \, , \quad   \gamma^{-1} Z_{ii} \gamma = e^{4 \pi i/3} \Z_{ii} \, ,
\end{align}
and trivially on all others.

This procedure is completely equivalent to the one decribed in \cref{app:skew_group}, with the slight advantage of making it easier to write down the superpotential for the orbifolded theory.

Each entry in \cref{eq:orbi_mat} is now a $3 \times 3$ matrix, thus the new theory will be a quiver gauge theory with 9 nodes, where the arrows are determined by the projections above. For example we have:
\begin{align}\label{eq:orbi_mat2}
    A_{22} \to \begin{pmatrix} A_{44} & 0 & 0 \\ 0 & A_{55} & 0 \\ 0 & 0 & A_{66}\end{pmatrix},  \quad X_{13} \to \begin{pmatrix} 0 & 0 & X_{19} \\ X_{27} & 0 & 0 \\ 0 & X_{38} & 0\end{pmatrix} \, ,
\end{align}
 and similarly for the other fields. As it can be explicitly checked, the resulting fields are the same ones one would get by applying the projection by $\gamma \in \mathbb{Z}_9$ with action $(1,1,2,6)$ on the fields $(A,X,Y,Z)$.

\providecommand{\href}[2]{#2}\begingroup\raggedright\endgroup

\end{document}